%% file: main.tex
\documentclass[ DIV=15 ]{scrartcl}  
 \pdfoutput=1 
\input{packages}
\input{pgfplotsettings}
\input{titleAndAuthors}

\begin{document}  
\normalem
\maketitle  
  
%% Abstract ---------------------------------------
%\vspace{-1.5cm} 
\hrule 
\section*{Abstract}
During the last ten years, increasing efforts were made to improve and simplify the process from Computer Aided Design (CAD) modeling to a numerical simulation. It has been shown that the transition from one model to another, i.e. the meshing, is a bottle-neck. Several approaches have been developed to overcome this time-consuming step, e.g. Isogeometric Analysis (IGA), which applies the shape functions used for the geometry description (typically B-Splines and NURBS) directly to the numerical analysis. In contrast to IGA, which deals with boundary represented models (B-Rep), our approach focuses on parametric volumetric models such as Constructive Solid Geometries (CSG). These models have several advantages, as their geometry description is inherently watertight and they provide a description of the models’ interior. To be able to use the explicit mathematical description of these models, we employ the Finite Cell Method (FCM). Herein, the only necessary input is a reliable statement whether an (integration-) point lies inside or outside of the geometric model. This paper mainly discusses such point-in-membership tests on various geometric objects like sweeps and lofts, as well as several geometric operations such as filleting or chamfering. We demonstrate that, based on the information of the construction method of these objects, the point-in-membership-test can be carried out efficiently and robustly.
 
%% Keywords ---------------------------------------
\vspace{0.25cm}
\noindent \textit{Keywords:} CSG, Constructive Solid Geometry, Finite Cell Method, Spline Ray Casting, Point Membership Classification, Procedural Modeling, Sweep, Loft
\vspace{0.25cm}
\hrule 

\vspace{0.25cm}
\noindent \copyright 2017. This manuscript version is made available under the CC-BY-NC-ND 4.0 license. \newline
Published in \textit{Computers \& Mathematics with Applications}\newline
https://www.sciencedirect.com/science/article/pii/S0898122117300780?via\%3Dihub \newline
DOI: 10.1016/j.camwa.2017.01.027 

\vspace{0.25cm}

\newpage
\tableofcontents
\newpage

%% Actual Content ---------------------------------
\input{introduction}
\input{overview}
\input{method}
\input{numericalExamples}

\input{conclusions}

\input{appendix}

\newpage
%% Acknowledgements -------------------------------  
\section*{Acknowledgements} 
The first and the last author gratefully acknowledge the
financial support of the German Research Foundation (DFG) under Grant RA 624/22-1.

%% References -------------------------------------
\bibliographystyle{ieeetr}
\bibliography{library}

\end{document}

%% file: packages.tex
\usepackage{color} 
\usepackage{psfrag} 
\usepackage{subfig}
\usepackage{hyperref}
\usepackage{amsmath}
\usepackage{graphics}
\usepackage{pstool}
\usepackage{todonotes}
          
\usepackage[square,sort,comma,numbers]{natbib}
 
\usepackage{pgfplots}
\usepackage{pgfplotstable}
\usepackage{filecontents}

\usepackage{booktabs}   
\usepackage{makecell} 
\usepackage{colortbl}   

\usepackage{xspace}
\usepackage{color}     
\usepackage{setspace}   
 
\usepackage{soul}
\usepackage{ulem}
\usepackage{currfile}
\usepackage{import}

\usepackage{authoraftertitle} % used for getting the authors 
\usepackage{authblk} % used for affiliations 

\usepackage{cleveref}
\usepackage{txfonts}
\usepackage{amsmath,amssymb}
\usepackage{float}
\usepackage{xfrac}
\usepackage{algorithm}
\usepackage[]{algpseudocode}
\usepackage{enumitem}   

%% file: pgfplotsettings.tex
\pgfplotsset{compat=1.8}

\newcommand{\findmax}[3]{
    \pgfplotstablesort[sort key={#2},sort cmp={float >}]{\sorted}{#1}%
    \pgfplotstablegetelem{0}{#2}\of{\sorted}%
    \let #3=\pgfplotsretval%
}

\pgfplotscreateplotcyclelist{myCycleListBlackAndWhite}
{%
  {black, mark=o},
  {black, mark=square},
  {black, mark=triangle},
  {black, mark=diamond}, 
  {black, mark=pentagon}, 
  {black, mark=*},
  {black, mark=square*},
  {black, mark=triangle*},
  {black, mark=diamond*}, 
  {black, mark=pentagon*}, 
}

\definecolor{darkgreen}{rgb}{0,0.4,0} 
\definecolor{darkbrown}{rgb}{0.5, 0.396, 0.09}

\definecolor{c1}{rgb}{0.0, 0.4196078431372549, 0.6431372549019608}
\definecolor{c2}{rgb}{1.0, 0.5019607843137255, 0.054901960784313725}
\definecolor{c3}{rgb}{0.6705882352941176, 0.6705882352941176,
0.6705882352941176} \definecolor{c}{rgb}{0.34901960784313724, 0.34901960784313724, 0.34901960784313724}
\definecolor{c4}{rgb}{0.37254901960784315, 0.6196078431372549,
0.8196078431372549} \definecolor{c}{rgb}{0.7843137254901961, 0.3215686274509804, 0.0}
\definecolor{c5}{rgb}{0.5372549019607843, 0.5372549019607843,
0.5372549019607843} \definecolor{c}{rgb}{0.6352941176470588, 0.7843137254901961, 0.9254901960784314}
\definecolor{c6}{rgb}{1.0, 0.7372549019607844, 0.4745098039215686}
\definecolor{c7}{rgb}{0.8117647058823529, 0.8117647058823529,
0.8117647058823529}

\pgfplotscreateplotcyclelist{myCycleListColor}
{%
  {blue, mark=o},
  {red, mark=square},
  {darkbrown, mark=triangle},
  {darkgreen, mark=diamond},
  {black, mark=pentagon},
  {blue, mark=*},
  {red, mark=square*},
  {darkbrown, mark=triangle*},
  {darkgreen, mark=diamond*},
  {black, mark=pentagon*},
}

\pgfplotsset{every axis/.append style= 
              {
                font=\small,
                mark size=2,
                line width = 0.1,
                legend style={font=\small, mark size=3, draw=none, fill=none},
                legend cell align=left,
                cycle list name=myCycleListColor,
              }
            }
            
\pgfplotstableset
{
  every odd row/.style={before row={\rowcolor[gray]{0.9}}},
  every head row/.style=
  {
    before row=
    {%
      \toprule
    },
    after row=\midrule
  },
  every last row/.style=
  {
    after row=\bottomrule
  }
}

\newif\ifdrawboundingbox

\usetikzlibrary{external} % Is activated with the command
\tikzset{external/system call={pdflatex \tikzexternalcheckshellescape
-halt-on-error -interaction=batchmode -jobname "\image" "\texsource"}} 
\pgfkeys
{ 
  /pgf/number format/.cd,
  fixed,
  set thousands separator={\,},
  1000 sep in fractionals,
}

%% file: titleAndAuthors.tex
\title{From geometric design to numerical analysis: A direct approach using the Finite Cell Method on Constructive Solid Geometry }
 \author[1]{Benjamin Wassermann \thanks{benjamin.wassermann@tum.de, Corresponding Author}}
\author[1]{Stefan Kollmannsberger\thanks{stefan.kollmannsberger@tum.de}}
\author[1]{Tino Bog\thanks{tino.bog@tum.de}}
\author[1,2]{Ernst Rank\thanks{ernst.rank@tum.de}}

\affil[1]{Chair for Computation in Engineering,
 Technische Universit\"at M\"unchen,
  Arcisstr. 21, 80333 M\"unchen, Germany}
  
\affil[2]{Institute for Advanced Study,
 Technische Universit\"at M\"unchen,
 Lichtenbergstr. 2a, 85748 Garching, Germany}

%% file: introduction.tex
\section{Introduction} \label{sec:intro}

Computer aided engineering in general requires an iterative process to find an optimal design. This iterative process consists of a modelling phase followed by a numerical simulation and an analysis phase. \\
Modern CAD tools mainly use two different techniques to create 3D models. A classic method, which is still commonly used, is boundary representation (B-Rep)\citep{Foley1997}. B-Rep describes a body implicitly as a topological model via its faces, edges, and nodes. Geometric information is then assigned to faces and edges, often using B-Spline-, or NURBS surfaces and curves. A more recent and natural approach is Procedural Modeling (PM), which is strongly related to Constructive Solid Geometry (CSG), but extends this concept by providing additional operations and primitives. Both CSG and PM describe a complex model as a combination of simple or complex primitives and Boolean operations (union, intersection, difference). Procedural modeling and B-Rep each have advantages and disadvantages, which are often complementary in such a way that, nowadays, many CAD systems use a hybrid representation combining B-Rep and PM~\citep{Gomes1991}. In this context, the B-Rep model provides information necessary e.g. for visualization purposes. PM serves as an underlying model that can easily be used for parametric and feature-based design~\citep{Shah1995}, for which a description of the construction history, the dependencies, and the constraints is mandatory. It is noteworthy that it is always possible to derive a B-Rep model from a PM model, but not the other way round. This is due to the loss of information in the conversion from PM to B-Rep. In addition, B-Rep cannot provide information about the structure of the interior of the model. However, this information can be crucial, for example in cases of heterogeneous materials or to describe additive manufacturing processes. Interestingly, fully three-dimensional-computational mechanical analyses mostly draw the geometrical information of the computational domain from B-Rep models which are then explicitly converted into a volumetric description by a meshing process. Moreover, in the finite element method (FEM), elements are required to conform with the physical boundaries of the model, which often requires a flawless B-Rep description. A practical consequence of these requirements is the often huge engineering effort to 'clean' a CAD model or to 'heal' a finite element mesh before a numerical analysis can start. At Sandia National Laboratories~\citep{Cottrell2009}, an estimation of the relative time required for a representative design process showed that more than 80 \% of the engineering effort is allotted to the transition from geometric models to simulation models that are suitable for analysis. 

Various methodologies have been developed to overcome the difficulties involved in this transition process. The most prominent method in the Computational Mechanics Community is the recently introduced Isogemoetric Analysis (IGA) as proposed by Hughes et al. \citep{Hughes2005}. IGA aims at bridging the gap between the CAD model and computational analysis by a closer mathematical interconnection between the two worlds. To this end, the same B-Splines and NURBS representations used to describe CAD are applied as both geometry and Ansatz functions in FEM. These functions offer several desirable properties such as the possibility of straightforward refinements in grid size and polynomial degree, as well as the possibility to control the continuity within a patch. Most importantly, they guarantee a precise description of the geometry, in contrast to classical FEM, where only an approximation can be obtained by meshing into tetrahedra or hexahedra. Furthermore, as B-Splines and NURBS are functions of higher order, they offer the potential to deliver high convergence rates if the underlying problem possesses smooth solutions. Concerning the modeling processes, IGA was first applied to B-Reps which consisted of several conforming two-dimensional B-Splines or NURBS patches. More complicated topologies are usually generated by trimming, which may lead to non-watertight geometric models. Remedies for this problem range from classic re-parametrization \citep{Piegl1997} to the use of T-Splines~\citep{Bazilevs2010a}. 

An alternative, designed to overcome the problems of B-Rep descriptions, are V-Reps, recently proposed by Gershon et al.~\citep{Massarwi2016}. They consist of trimmed trivariate NURBS patches which directly describe the volume under consideration. A related approach was presented by Zuo et al. \citep{Zuo2015}, who proposed to treat CSG primitives separately as volumes using IGA and to trim and glue them by using the Mortar Method ~\citep{Patera1988} at their intersection surface. However, apart from some special numerical pitfalls inherent to domain sewing techniques, this poses the additional difficulty that an explicit boundary representation needs to be set up \emph{for all} inter-subdomain boundaries. Another related approach was presented much earlier by Natekar et al.~\cite{Natekar2004} who proposed to combine spline-based element formulations with two-dimensional CSG model descriptions. However, this approach is also based on heavy use of explicit boundary representations as well as a decomposition into sub-domains. The same holds also for the design-through analysis procedure presented in~\cite{Schillinger2012a}, which uses a B-Rep description and relies on a 3D ray-casting test to describe the volume of the model.

This strong reliance on the explicit description of coupling interfaces -- or, more generally, surface descriptions in the analysis process -- poses a drawback to parametric modeling approaches: Even though a change of parameters or constraints hardly has any impact on the general structure of the CSG model itself, it often triggers a complete reconstruction of the entire corresponding B-Rep model. Together with the observation that a CSG or a procedural model is intrinsically watertight and directly provides information about the interior, we conclude that a desirable simulation technique would have to use the explicit description of volumes by CSG as often as possible, and its B-Rep representation as little as possible. 

To this end, we propose a combination of CSG and the Finite Cell Method for volume orientated modeling and numerical analysis. We denote this approach as a direct modeling-to-analysis method as it allows, like IGA, a very close interaction of the (geometric) design process and the (numerical) analysis, where an engineer can immediately investigate consequences of a variation of the geometric design on the mechanical behavior of a structural object. \\
The Finite Cell Method (FCM)~\citep{Parvizian2011}, which represents the core of this approach, is a high-order fictitious domain method that embeds an arbitrary complex geometry into an extended domain which can easily be meshed by a Cartesian grid. The complexity of the geometry is handled only on the level of integration of element matrices and load vectors. This makes the method very flexible, because the only information the FCM needs from the CAD model is a reliable and robust point-in-membership test, i.e. whether an integration point lies inside or outside of the physical model. This point-in-membership test is directly provided by the CSG model description. The interplay between CSG and FCM was already investigated for simple primitives, and it proved to be an accurate and efficient method to analyze trimmed NURBS patch structures ~\citep{Rank2012}. The goal of the present paper is to extend the combination of the FCM and the CSG to more complex geometric models as well as to solid construction processes of industrial relevance. \\
This paper is organized as follows: In~\cref{sec:overview}, a short overview on geometric representations and the Finite Cell Method is given. In~\cref{sec:method}, the relevant methods for the combination of CSG and FCM are presented. \Cref{sec:numExp} provides examples showing the relevance and potential for practical applications before conclusions are drawn in~\cref{sec:conclusions}.
\newpage

%% file: overview.tex
\section{From Geometric Design to Numerical Analysis} \label{sec:overview}
\subsection{Geometric modelling}
In the field of Computer Aided Design (CAD), several different schemes are available to model 3D geometric objects. Nowadays, 3D CAD systems are usually based on either (i) boundary representation or (ii) procedural modeling with solid primitives. Next, the two schemes will be outlined -- followed by a short section about the conversion of one into another. 

\subsubsection{Boundary Representation}
\begin{figure}[H]	
	\centering
	\includegraphics[width=12cm]{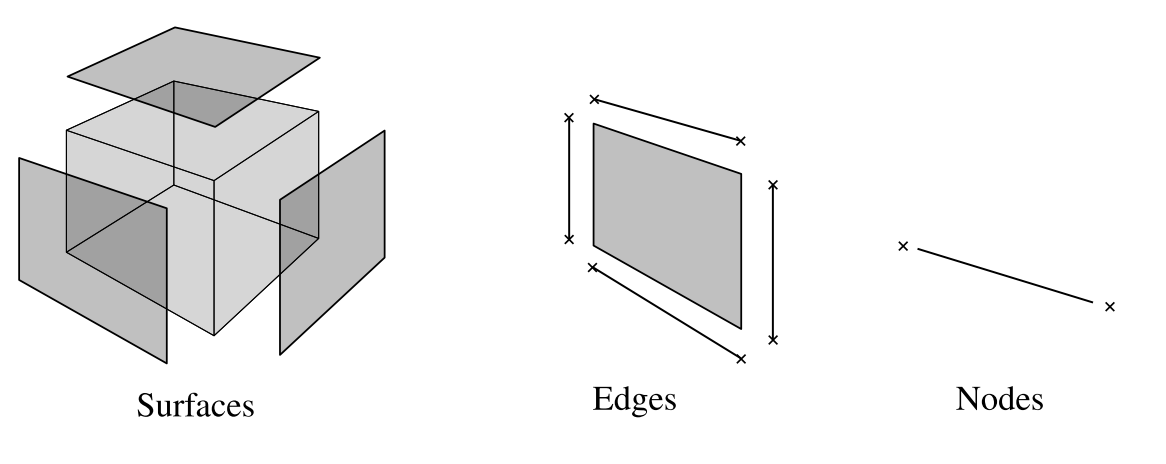}
	\caption{Structure of B-Rep.}
	\label{fig:BRep}	
\end{figure}

Classically, objects are defined by a Boundary Representation (B-Rep), where only the objects' surfaces with their corresponding edges and nodes are stored (see \cref{fig:BRep}) \citep{Bungartz2004}. This is motivated by the requirements for visualization, in the scope of which 3D objects are displayed via their surfaces. 

Although B-Rep has several advantages, for example the direct access to surfaces, it has also some disadvantages, especially with respect to a subsequent numerical simulation. B-Rep models are not necessarily watertight, which means that a point-in-membership test on these potentially corrupted solids may not be sufficient to clearly distinguish whether a point lies outside or inside the domain.

Another disadvantage are defective topological descriptions, such as multiple nodes, or edges. Although such errors do not disturb the visualization, they may render numerical simulations difficult or even impossible. These 'dirty geometries' are among to the major reasons for the considerable effort that often goes with cleaning up a geometric model. This preparation work is necessary to be able to mesh a model into a consistent finite element model.

\subsubsection{CSG and procedural modelling} \label{sec:CSGModelling}
\begin{figure}[H]	
	\centering
	\includegraphics[width=10cm]{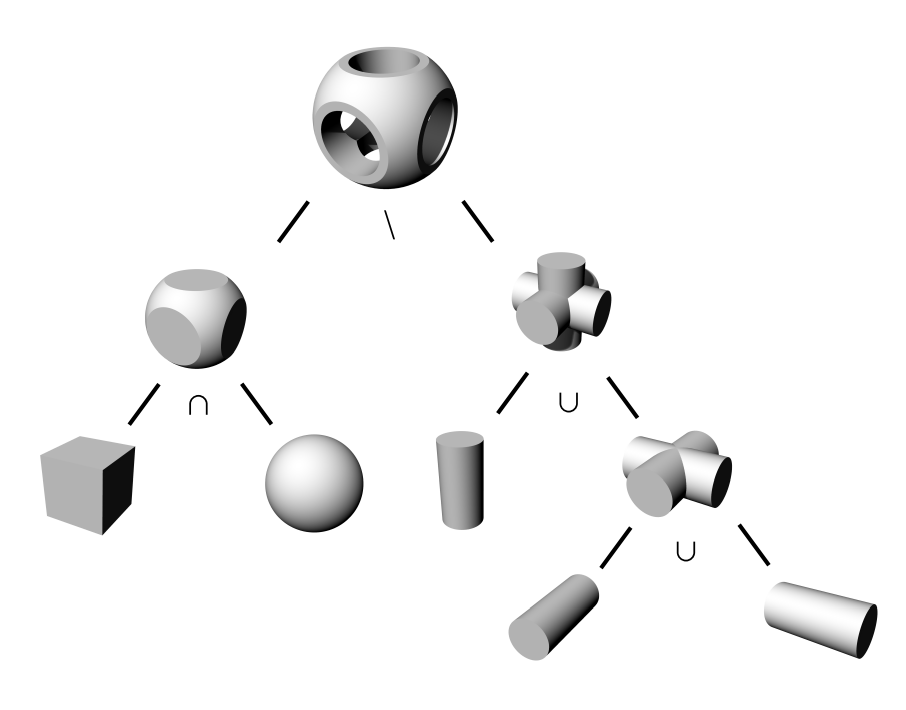}
	\caption{CSG Tree using the three boolean operations: union $\cup$, intersection $\cap$, difference $\backslash$ on primitives.}
	\label{fig:CSG}	
\end{figure}

Alternatively, a 3D object can be described as a procedural model that is strongly related to Constructive Solid Geometry (CSG) \citep{Requicha1977}. In CSG, a 3D object is created from a set of primitives, such as cubes, cylinders, cones, spheres etc. These primitives are combined by the three basic boolean operations: union, intersection, and difference. The resulting CSG object is stored implicitly in a CSG tree (see \cref{fig:CSG}  ).

\begin{figure}[H]
	\centering
	\includegraphics[width=10cm]{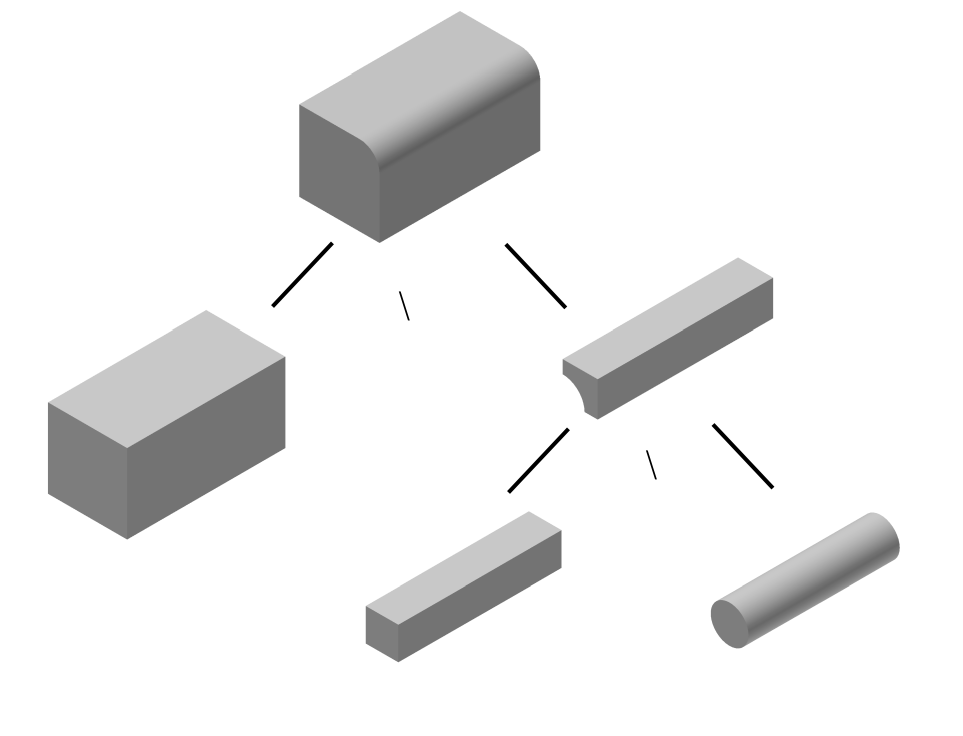}
	\caption{ Extended operations can be expressed by the classical boolean operations: union, intersection, difference. The example shows filleting an edge.}
	\label{fig:ExtendedOperation}	
\end{figure}

In contrast to CSG, a procedural model stores the construction steps in chronological order as a construction history.
 These two different storage schemes, the CSG tree and the construction history, can be converted one to another. Procedural modeling also comes along with a richer set of operations and primitives, in the following referred to as extended operations and extended primitives. Extended operations include chamfer, fillet, drilling a hole, and draft. A closer look reveals that they are in fact just a sequence of the original three boolean operations -- union, difference, and intersection -- which are summarized for convenience (see \cref{fig:ExtendedOperation}).

\begin{figure}[H]
	\centering
	\includegraphics[width=14cm]{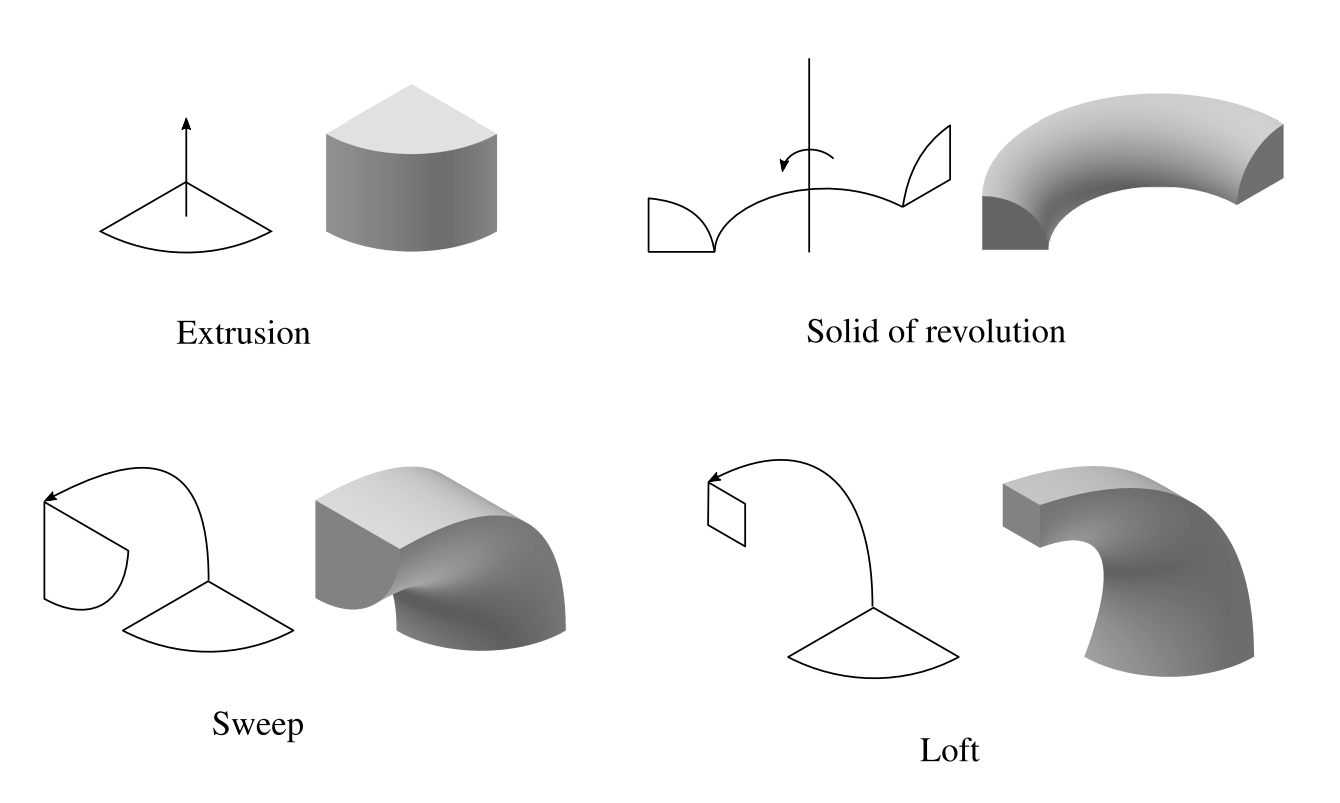}
	\caption{Extended primitives: extrusion, solid of revolution, sweep, loft.}
	\label{fig:ExtendedPrimitives}	
\end{figure}

However, extended primitives such as extrusions, sweeps, lofts, and solids of revolution can be regarded as a true extension to the CSG primitive set (see \cref{fig:ExtendedPrimitives}). 
 
\noindent The construction processes of these primitives are all strongly related. On an arbitrary plane, a 2-dimensional closed contour line is drawn. This is extruded along a sweep path. Depending on the shape of this path, either an extrusion, a solid of revolution, or a sweep is obtained. Only the  construction of a loft differs slightly. Here, the initial sketch is not just extruded, but is blended along the path into another (final) sketch. 

Regarding the aspect of numerical simulation, one big advantage of the procedural model over the B-Rep model is that it is inherently watertight. All primitives are explicitly described by their volume and always form a valid closed 3D object. CSG operations combine these valid primitives, and their combination again results in a valid model. 

\subsubsection{Conversion between explicit and implicit models} \label{sec:surfaceRecovery}
We would like to stress the fact that a conversion from a B-Rep model to a CSG model is usually not possible, as B-Rep models carry less information. Thus, it is desirable to use an explicit volume description such as CSG directly for a simulation.

The drawback of the CSG description is that it only provides indirect access to the models' surfaces, which might be needed e.g. for visualization. Fortunately, it is always possible to derive an approximate surface description from the CSG, e.g. by the marching cubes algorithm \citep{Lorensen1987}. This algorithm only needs the information whether a point is inside or outside of the model at any point, which is readily available via the CSG construction tree. Another (practical) possibility is to gather the B-Rep information directly from the CAD software if needed. This is often possible because many CAD systems maintain a B-Rep model concurrently to a CSG model and provide direct access to its surface. These derived surface models, even if they are not perfectly watertight, are a sufficient basis to impose boundary conditions in an analysis by the Finite Cell Method, which will be described in the next section.

\subsection{Finite Cell Method}
The finite cell method (FCM) is a fictitious domain approach using high order finite elements. It relies on an explicit description of the volume of the physical domain and is able to deliver high accuracy~\citep{Duster2008}.

\subsubsection{Classical finite elements}
Consider a linear-elastic problem on a physical domain $\Omega_{\mathrm{phy}} $ with the boundary $\delta\Omega$ divided into Dirichlet and Neumann boundary parts $\Gamma_{\mathrm{D}}$ and $\Gamma_{\mathrm{N}}$. By applying the principle of virtual work, the weak form of an elliptic partial differential equation reads \citep{Hughes2000}

\begin{equation}
	\mathcal{B}(\mathbf{u},\mathbf{v}) = \mathcal{F}(\mathbf{v}) \quad\quad\quad 
	\begin{matrix} 
	\forall \mathbf{u} \in \mathcal{S}(\Omega_{\mathrm{phy}} ) \\	 
	\forall \mathbf{v} \in \mathcal{V}(\Omega_{\mathrm{phy}} ) 
	\end{matrix} 
\end{equation}
with
\begin{equation}
	\mathcal{B}(\mathbf{u},\mathbf{v}) =  \int_{\Omega_{\mathrm{phy}} } \nabla \mathbf{v} : \mathbb{C} : \nabla \mathbf{u}\; \text{d}\Omega_{\mathrm{phy}}
\end{equation}
\begin{equation}
	\mathcal{F}(\mathbf{v}) =  \int_{\Omega_{\mathrm{phy}} } \mathbf{b} \cdot \mathbf{v}\; \text{d}\Omega_{\mathrm{phy}}  + \int_{\Gamma_{\mathrm{N}}} \mathbf{\hat{t}} \cdot \mathbf{v}\; \text{d}\Gamma_{\mathrm{N}}
	\label{eq:linFunctional}
\end{equation}
where $\mathbf{u}$ is the displacement, $\mathbf{v}$ the test function, and $\mathbb{C}$ the elasticity tensor. $\mathbf{b}$ and $\mathbf{\hat{t}}$ denote the body load and the prescribed boundary
traction applied on the Neumann boundary, respectively.  $\mathcal{S}(\Omega_{\mathrm{phy}} )$ is the trial function space, which is constructed such that $\mathbf{u}$ satisfies the prescribed Dirichlet boundary conditions $\mathbf{\hat{u}}$

\begin{equation}
 \mathcal{S} (\Omega_{phy} ) = \left\lbrace \mathbf{u}\;\;|\;\;\mathbf{u}  \in H^1(\Omega_{\mathrm{phy}} ), \mathbf{u} =\mathbf{\hat{u}} \;\; \forall\;  \mathbf{x} \in \Gamma_{\mathrm{D}} \right\rbrace ,
\end{equation}
whereas $\mathcal{V}(\Omega_{\mathrm{phy}})$ denotes the space of all admissible test functions that satisfy homogeneous Dirichlet boundary conditions
 \begin{equation}
 \mathcal{V} (\Omega_{\mathrm{phy}} ) = \left\lbrace \mathbf{v} \;\;|\;\;\mathbf{v}  \in H^1(\Omega_{phy} ), \mathbf{v} = 0 \;\; \forall\;  \mathbf{x} \in \Gamma_{\mathrm{D}} \right\rbrace .
\end{equation}
$H^1$ denotes the Sobolev space \citep{Reddy1997} of first order. Both $\mathbf{u}$ and $\mathbf{v}$ are discretized to yield an approximate solution using a linear combination of Ansatz functions $ \lbrace N_1, N_2, . . . , N_n \rbrace $

\begin{equation}
 	\mathbf{u}_h = \mathbf{\tilde{N}}\mathbf{u} , 
\end{equation}
where $\tilde{u}_i$ is the degree of freedom of the related Ansatz function $N_i$. Following the Bubnov - Galerkin approach \citep{Hughes1978}, the test functions $\boldsymbol{v}$ are represented by the same basis as the trial functions $\mathbf{v} \in \mathcal{S}(\Omega_{\mathrm{phy}} )$. This approach leads to the following system of linear equations:

\begin{equation}
 	\mathbf{\tilde{K}} \mathbf{u}= \mathbf{f}
\end{equation}
where $\mathbf{K}$ is the stiffness matrix and $\mathbf{f}$ the load vector.

\subsubsection{Concept of FCM}
\begin{figure}[h]
	\centering
	\includegraphics[width=16cm]{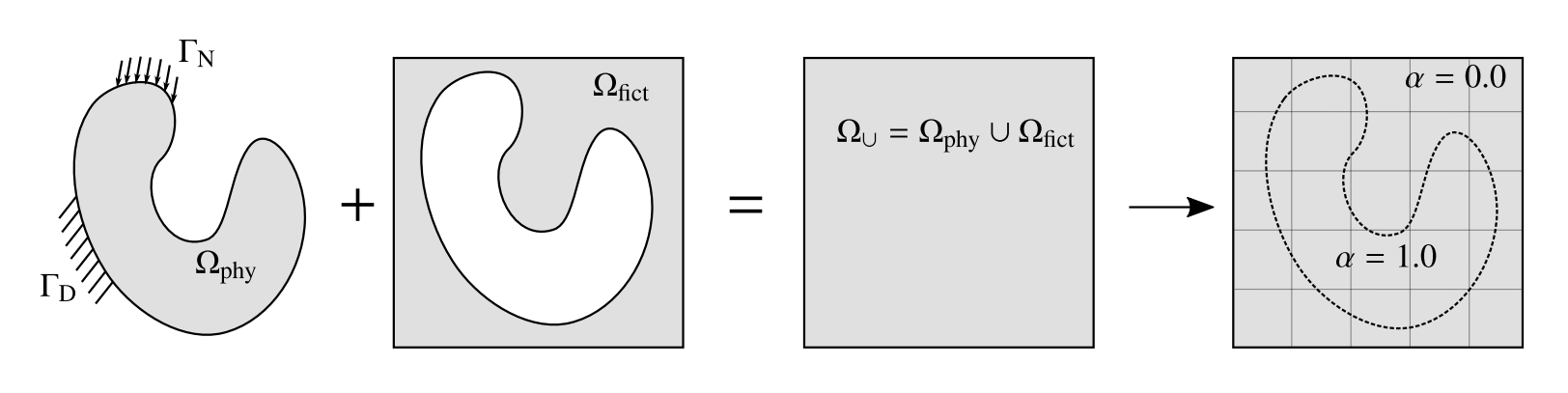}
	\caption{ Concept of Finite Cell Method.}
	\label{fig:FCM}	
\end{figure}

The original idea of the finite cell method is to extend the physical domain $\Omega_{\mathrm{phy}}$ by a fictitious domain $\Omega_{\mathrm{fict}}$ such that the resulting domain $\Omega_{\cup}$ has a simple shape, which can be meshed easily (see \cref{fig:FCM})\citep{Parvizian2007, Duster2008}.

The weak formulation is modified by extending integrals over the domain $\Omega_{\cup}$. Additionally, the virtual work terms are multiplied by a scalar field $\alpha(x)$:

\begin{equation}
	\mathcal{B}_e(\mathbf{u},\mathbf{v}) =  \int_{\Omega_{\cup}} \nabla \mathbf{v} : \alpha\mathbb{C} : \nabla \mathbf{u}\; d\Omega_{\cup} \label{eq:bilinearForm}
\end{equation}
\begin{equation}
	\mathcal{F}_e(\mathbf{v}) =  \int_{\Omega_{\cup}} \alpha\mathbf{b} \cdot \mathbf{v}\; d\Omega_{\cup} + \int_{\Gamma_N} \mathbf{\hat{t}} \cdot \mathbf{v}\; \Gamma_N 
\end{equation}

with $\alpha$ defined as:

\begin{equation}
\alpha = 
\left\lbrace \begin{matrix}
	1  \\
	10^{-q}
\end{matrix}
\right. \quad \quad
\begin{array}{l}
	\forall \mathbf{x} \in \Omega_{\mathrm{phy}} \\
	\forall \mathbf{x} \in \Omega_{\mathrm{fict}}
\end{array} \quad ,
\end{equation}
where, ideally $q\rightarrow \infty$. In practical applications, it is usually sufficient to choose $q=8 \;to \;10$. In essence, the discontinuous indicator function $\alpha$ now represents the geometric description of the domain. The convergence of this scheme is mathematically proven in~\cite{Dauge2015a} where it is also shown that the influence of a non-infinite $q$ is proportional to a (controllable) modeling error. 

After discretizing the extended domain $\Omega_{\cup}$ into a Cartesian grid, high-order finite elements can be used for the computation of the displacement field. Several different Ansatz functions for high-order elements are available, such as integrated Legendre polynomials \citep{Szabo2004, Parvizian2007}, B-Splines \citep{Schillinger2011, Schillinger2012b}, and Lagrange polynomials \citep{Duczek2013, Joulaian2013}.

\begin{figure}[H]
	\centering
	\includegraphics[width=12cm]{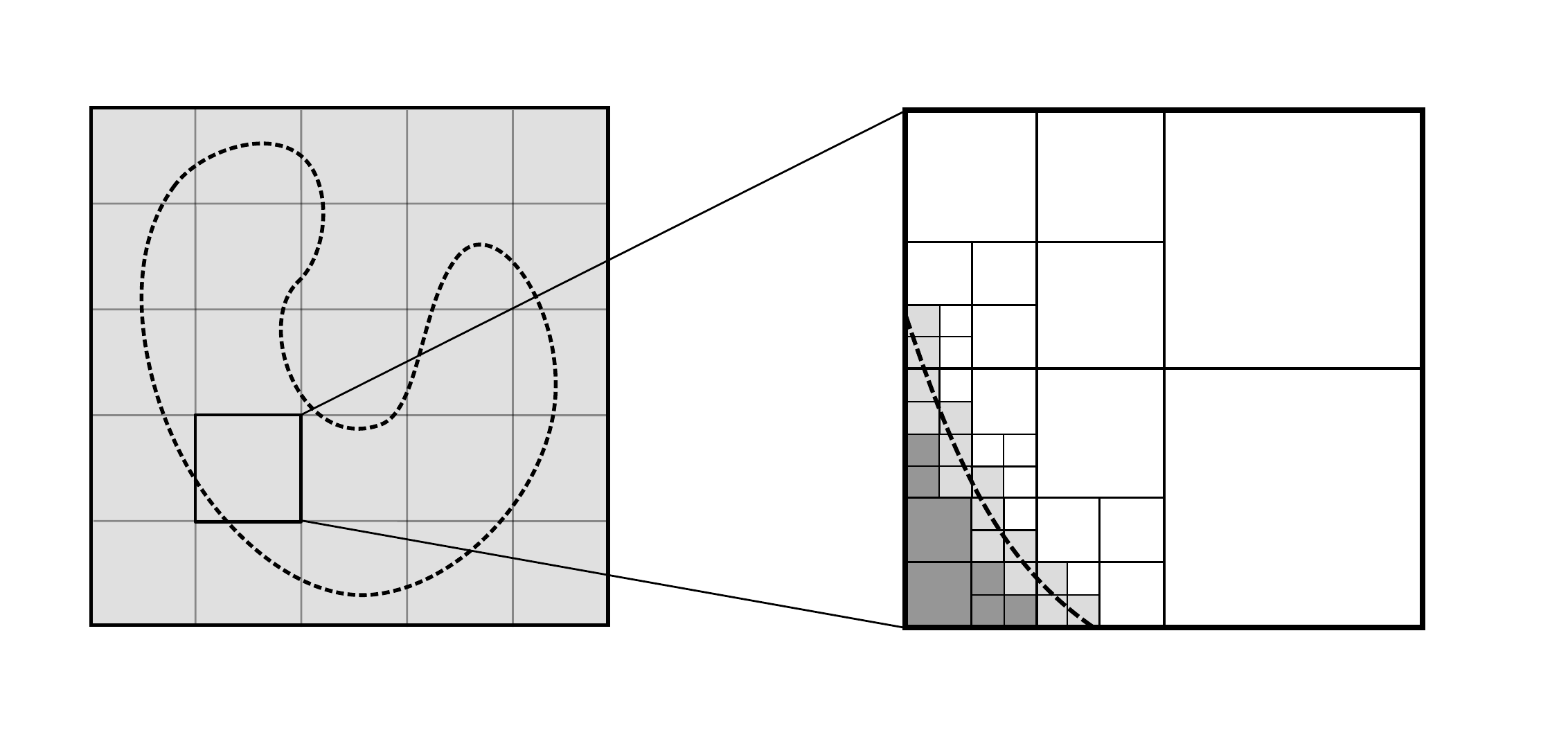}
	\caption{Quadtree partitioning of a 2D cell with a partitioning depth of 4. Coloring of quadtree leaves: white -- completely inside, light  gray -- cut, dark gray -- completely outside)}
	\label{fig:octree}
\end{figure}

The discontinuity of $\alpha$ necessitates an adaptive integration of the element matrices and load vectors, see e.g.~\cite{Abedian2013,Kudela2016} for a recent overview of possible schemes. The simplest (although not most efficient) choice is a composed integration by means of an octree in 3D or a quadtree in 2D (see \cref{fig:octree}) which is used in all examples presented in~\cref{sec:numExp}.

\subsubsection{Boundary conditions}
In FCM the boundaries of the physical domain $\Omega_{\mathrm{phy}}$ typically do not coincide with the boundaries of the cells in the extended domain $\Omega_{\cup}$. In that case boundary conditions are enforced weakly.\\
Homogeneous Neumann boundary conditions need not to be treated in a special manner since they are naturally included in the weak formulation (see equation \eqref{eq:bilinearForm}). Inhomogeneous Neumann conditions can be handled by integrating the prescribed traction forces $\mathbf{\hat{t}}$ along the boundary $\Gamma_{\mathrm{N}}$ (see equation \eqref{eq:linFunctional}). \\
Dirichlet boundary conditions, i.e. applying a prescribed displacement $ \mathbf{u}= \mathbf{\hat{u}} \;\; \forall \mathbf{x} \in \Gamma_{\mathrm{D}}$, must be applied in a weak sense using, e.g., Lagrangian multipliers, the penalty method, or Nitsche's method, see e.g. \cite{Kollmannsberger2015} \cite{Ruess2013}. 

In short, an explicit description of the boundary must only be available where forces or displacements are imposed (see \cref{sec:surfaceRecovery}). The geometric and topological requirements of these surfaces are much lower than for mesh-generation as they only need to fulfill conditions for a sufficiently accurate numerical integration. Therefore, these integration meshes neither have to be conforming or watertight, as it would be necessary for the basis for a finite element computation.  

%% file: method.tex
\section{FCM and CSG} \label{sec:method}

\subsection{Point-in-membership test}
\begin{figure}[H]\centering
	\includegraphics[width=10cm]{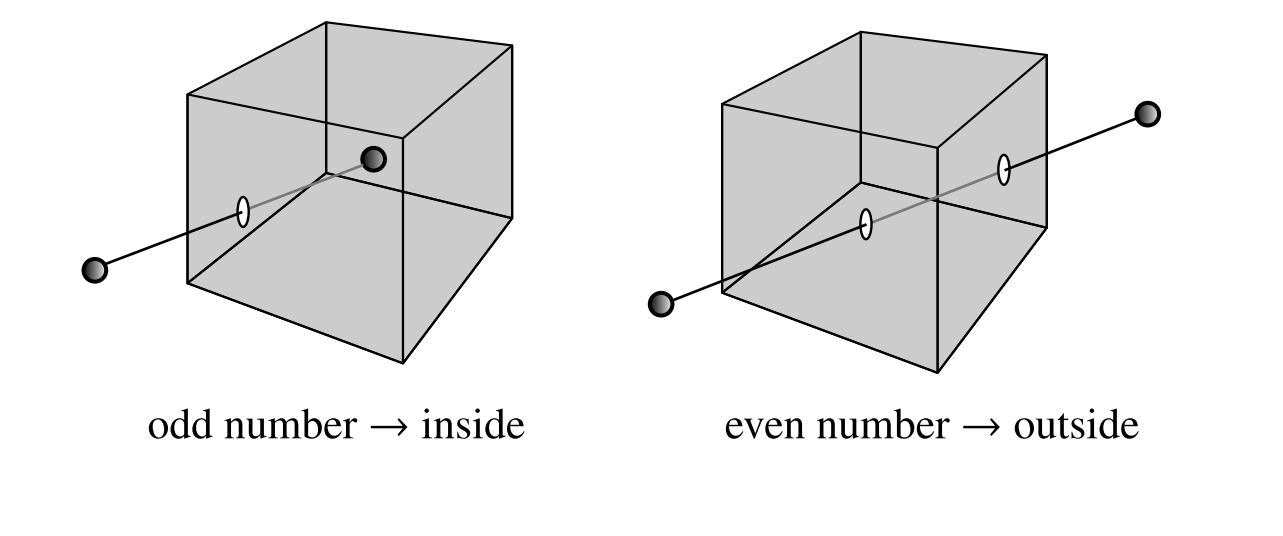}
	\caption{ Ray casting on a B-Rep model.}
	\label{fig:RayCastBRep}	
\end{figure}

The geometric description of the physical domain is provided by the function $\alpha$ which is explicitly given by a point-in-membership test, i.e. if a point lies inside the physical domain $\Omega_{\mathrm{phy}}$ or inside the fictitious domain $\Omega_{\mathrm{fict}}$. This test can be carried out on a B-Rep model by ray casting. To this end, a ray is sent out from the integration point into an arbitrary direction, and all intersections with the surfaces are counted. If the number of intersections is odd, then the starting point lies inside $\Omega_{\mathrm{phy}}$, otherwise it lies in $\Omega_{\mathrm{fict}}$ (see \cref{fig:RayCastBRep}). This simple test, however, can fail for non-water-tight models. Moreover, this approach can become quite expensive for highly resolved models or models with a complex surface. Even though there are methods available to reduce the amount of ray castings. \footnote{For example hierarchical bounding boxes, which divide the surface triangles into small chunks. Only if the ray intersects a bounding box the query is forwarded to the respective next level of bounding boxes and finally to a the relevant chunk(s) of triangles.}

\begin{figure}[H]\centering
	\includegraphics[width=14cm]{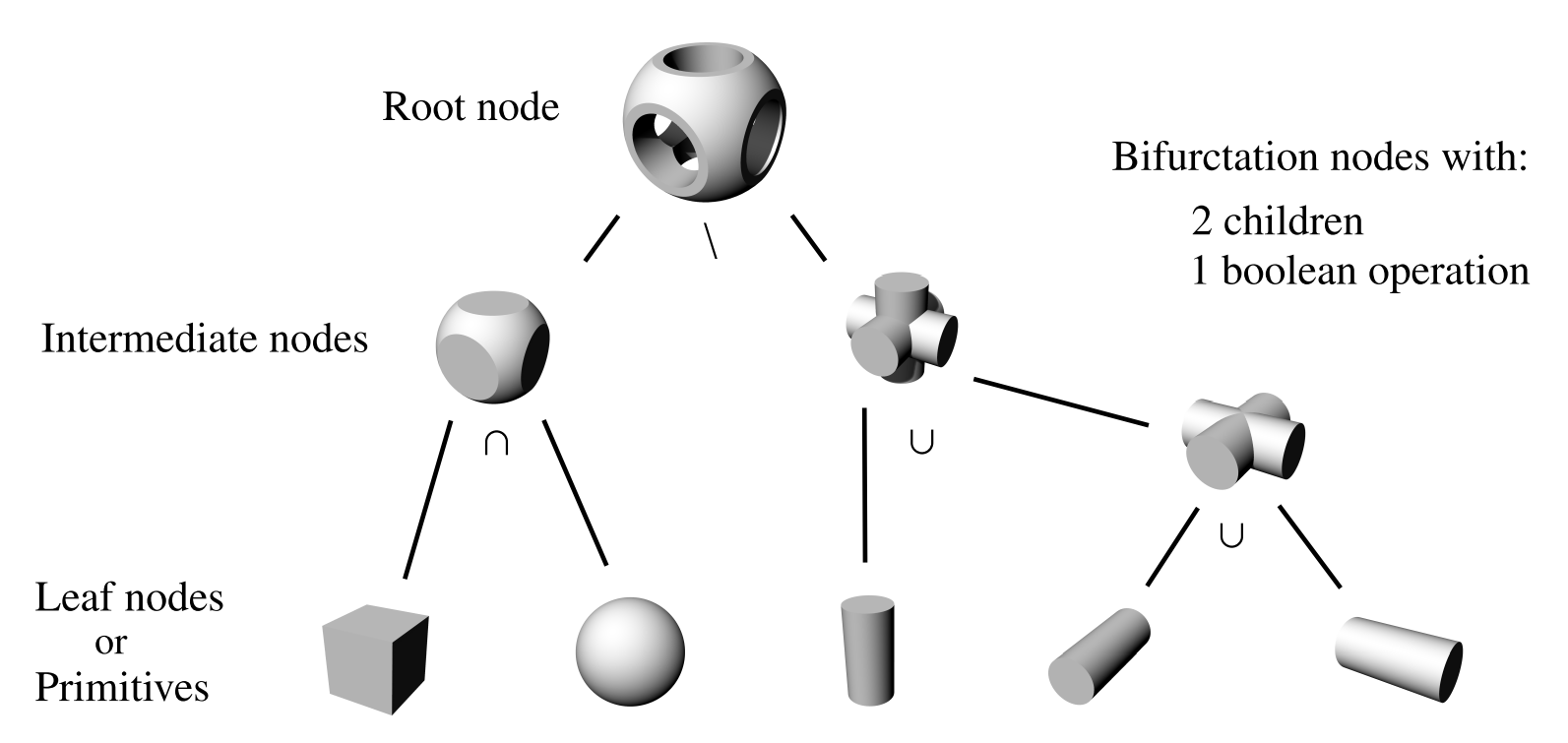}
	\caption{ CSG-tree consisting of (a) primitives at leaf nodes and (b) bifurcation nodes at an intermediate or root node with two children and a boolean operation.}
	\label{fig:csgTree}	
\end{figure}

In contrast, a point-in-membership test can be performed much faster on a CSG tree of classical primitives. In our implementation the CSG tree is a full binary-tree, i.e. each node has either exactly two or zero child nodes. There are two different types of nodes: (a) primitives which are always leaf nodes, and (b) bifurcation nodes which combine their two children with one of the three boolean operations: intersection, union, difference (see \cref{fig:csgTree}). It is also possible to add elements, carrying only one child like unary operations such as negations, or transformations to the CSG tree. Whereas negations have hardly any application in geometric modeling of solid mechanics (in contrast to exterior problems, such as fluid mechanics), transformations are used intensively. To this end, simple transformations, such as translation, scaling and rotation can be handled directly on the tree nodes, whereas more complex operations, like mirroring, would be preferably represented by a node, containing the operation and only the respective child to which this operation should be applied. \\
For a point-in-membership test, first the root node representing the final construction is queried. In the unlikely case that the root node is also a leaf node and, hence, a primitive, this test is carried out directly. In all other cases, the root element is a bifurcation node. Then, the request is forwarded recursively to its children until it reaches a primitive, i.e. a leaf node. The pairing leaf node is tested as well. For simple primitives (treated in \cref{sec:PIM_simple}), this test can be carried out very fast, as an analytical solution is available. Both results are then combined with the logical operation defined by the parent bifurcation node. \\
The algorithm can be sped up considerably by the following considerations: Due to the recursive property at a bifurcation node, the entire branch of the first child is evaluated before the second child is queried. This can be used with the knowledge that, in case of an intersection or a difference, it is often not necessary to test the entire tree. Considering a difference
\begin{equation}
	 A \backslash B := \mathbf{P} \in A \wedge \mathbf{P} \notin B 
\end{equation}
i.e. if point $\mathbf{P}$ is not in body A, body B needs not to be queried. Hence, entire branches of the CSG tree can be omitted during the query. The same holds for the intersection. Therefore, it is useful to perform an intersection test with the bounding box of computationally expensive branches first. This test can be introduced on each level in the CSG tree.

\subsection{Point-in-membership test on simple primitives}\label{sec:PIM_simple}
For classical primitives, a simple analytical description is available. Hence, a point-in-membership test can be carried out very efficiently. Consider a primitive $\mathbf{B}_i$ which is created axis-aligned on the $x-y$ plane, and assume that we define each primitive as a closed body, i.e. the boundary is included in the body. The test whether a point of interest $\mathbf{P} = \{x,y,z\}$ is inside a primitive reads as follows for a:

\begin{itemize}
	\item \textbf{Sphere} with center point $\mathbf{C}_{\text{Shpere}}$ and radius $r_0$
	\begin{equation}
		\mathbf{P} \in \mathbf{B}_{\text{Sphere}} \quad \text{iff} \quad \quad  ||\overline{\mathbf{PC}}_{\text{Sphere}}||_2  \leq r_0,
	\end{equation}
		
  \item for a \textbf{Cuboid} defined by two corner points lying on its diagonal $P_{\text{start}} = [x_s,y_s,z_s]$ and $P_{\text{end}} = [x_e,y_e,z_e]$	
		\begin{equation}
		\mathbf{P} \in \mathbf{B}_{\text{Cuboid}} \quad \text{iff} \quad \quad  x \in [x_s, x_e] \wedge  y \in [y_s, y_e] \wedge z \in [z_s, z_e],
	\end{equation}
	
	\item for a \textbf{Cylinder} defined by its center point $\mathbf{C}_{\text{Cylinder}} = \{x_c, y_c, z_c \equiv 0\}$, radius $r_0$, and height $h_0$ 
	\begin{equation} 
		\mathbf{P} \in \mathbf{B}_{\text{Cylinder}} \quad \text{iff} \quad \quad   ||\overline{\mathbf{\tilde{P}C}}_{\text{Cylinder}}||_2  \leq r_0 \wedge z \in [z_c , z_c + h_0]
	\end{equation}	
where point $\mathbf{\tilde{P}} = \{x,y,0\}$ is the projection of point $\mathbf{P}$ onto the $x-y$ plane,

	\item a \textbf{Cone Frustum} with the same set up as for the cylinder whose bottom and tip circles are concentric with radii $r_0$ and $r_1$ 
	\begin{equation} 
		\mathbf{P} \in \mathbf{B}_{\text{Cone}} \quad \text{iff} \quad \quad   ||\overline{\mathbf{\tilde{P}C}}_{\text{Cone}}||_2  \leq r(z) \wedge z \in [z_c , z_c + h_0]
	\end{equation}
	with
	\begin{equation} 
		r(z) = \frac{r_1 -r_0}{h_0}  z + r_0.
	\end{equation}
	If the radius $r_1$ is chosen to be zero, a complete cone is obtained.
	
	\item and a \textbf{Pyramid Frustum} with a corresponding set up to that of  the cone frustum. The rectangular bounding box at the bottom $[x_{s0}, x_{e0}]$,  $[y_{s0}, y_{e0}]$ and on the top $[x_{s1}, x_{e1}]$,  $[y_{s1}, y_{e1}]$ have the same center point.
	\begin{equation}
		\mathbf{P} \in \mathbf{B}_{\text{Pyramid}} \quad \text{iff} \quad \quad  x \in [x_s(z), x_e(z)] \wedge  y \in [y_s(z), y_e(z)] \wedge z \in [z_c , z_c + h_0]
	\end{equation}
	with
		\begin{align}
			x_s(z) = \frac{x_{s1} - x_{s0}}{h_0}  z + x_{s0} 
		\end{align}	
			
	and $ x_e(z), y_s(z), y_e(z)$ correspondingly. If $x_{s1} = x_{e1}$ and $y_{s1} = y_{e1}$ the pyramid frustum becomes a complete pyramid.
\end{itemize}

\noindent There are also fast analytical solutions for other primitives –- such as wedges, four-sided pyramids, or tori -- available.\\
In general these primitives are not constructed axis-aligned to the $x-y$ plane.  At a suitable position, a local orthonormal coordinate system $\mathbf{A}\left(\mathbf{A}_1, \mathbf{A}_2, \mathbf{A}_3\right)$ is thus constructed, where $\mathbf{A}_i$ denotes the respective base vectors. The oriented primitive can be constructed on the local $\mathbf{A}_1$-$\mathbf{A}_2$ plane,. To perform a point-in-membership test, the point of interest $\mathbf{P}$ needs to be mapped from the Cartesian coordinate system $\mathbf{E} \left(\mathbf{E}_1,\mathbf{E}_2,\mathbf{E}_3\right)$ to the local base $\mathbf{A} $. 

	\begin{equation} 
		\mathbf{\tilde{P}}= \mathbf{T} \:\mathbf{P} + \mathbf{v}
		\label{eq:PointMapping}
	\end{equation}	
with $\mathbf{v}$ the translation vector between the origins of the Cartesian $\mathbf{E}$ and local basis system $\mathbf{A}$
	\begin{equation} 
		\mathbf{v} = \mathbf{O}_A - \mathbf{O}_E = \mathbf{O}_A
	\end{equation}	
and the transformation matrix
	\begin{equation} 
		 \mathbf{T} = \begin{bmatrix}
		 A_{1x} & A_{2x}  & A_{3x}\\
		 A_{1y} & A_{2y}  & A_{3y}\\
		 A_{1z} & A_{2z}  & A_{3z}
		 \end{bmatrix} \quad \quad \quad .
	\end{equation}

\subsection{Point-in-membership test on extended primitives} \label{sec:PIM_Extended}
Point-in-membership tests are more complex, if the primitives are generated by sweeps or lofts, where no analytical tests are available in general. Nevertheless, it is possible to perform a fast, reliable test on these geometries as well. The basic idea is to reduce the dimension of the problem, taking advantage of the fact that these extended primitives are constructed by moving 2D sketches along a curve.We remark that these sketches are planar which corresponds to the options available in common CSG modeling tools such as Autodesk\textsuperscript{\textregistered} Inventor\textsuperscript{\textregistered} and Siemens NX\textsuperscript{\textregistered}. The fallback to two dimensional point in membership tests does not compromise robustness because it is much easier to construct such tests in two- than in three dimensions. In the subsequent sections we present a fast and robust point in membership test based on ray casting. Other robust alternatives such as a classic winding number test~\citep{Whitney1937} are possible as well.

\subsubsection{Coordinate systems}
In the following two local basis systems are introduced. First, the local basis $\mathbf{A}(\xi)$, which belongs to the curve, e.g. the Frenet base, and changes according to the (sweep or loft) path variable $\xi$. Second, the local basis $\mathbf{B}(\xi)$ of the sketch. This basis $\mathbf{B}(\xi)$ and so the sketch change accordingly to the curve basis system $\mathbf{A}(\xi)$ and thus depends also on $\xi$.
\begin{subequations}
\begin{equation}
	\mathbf{A}(\xi)
	\begin{pmatrix}
		\mathbf{A}_1(\xi), & \mathbf{A}_2(\xi), & \mathbf{A}_3(\xi)
	\end{pmatrix}
\end{equation}
\begin{equation}
	\mathbf{B}(\xi) 
	\begin{pmatrix}
		\mathbf{B}_1(\xi), & \mathbf{B}_2(\xi), & \mathbf{B}_3(\xi)
	\end{pmatrix}
\end{equation}
\end{subequations}

\begin{figure}[H]\centering
	\includegraphics[width=12cm]{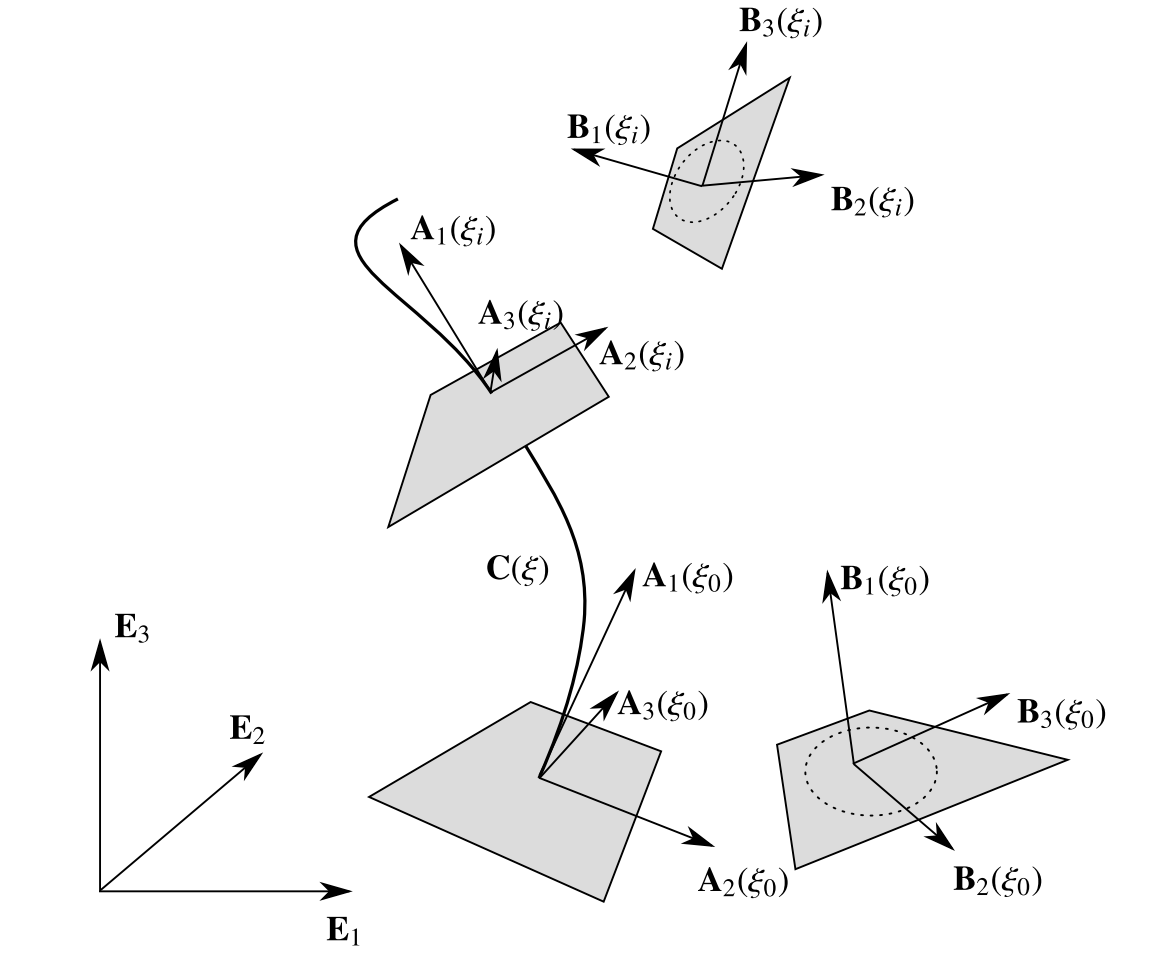}
	\caption{ Coordinate systems: (a) Cartesian basis $\mathbf{E}$. (b) Local coordinate system $\mathbf{A}(\xi)$ of the sweep path  $\mathbf{C(\xi)}$ \\(c) Local coordinate system of the sketch $\mathbf{B}(\xi)$.  }
	\label{fig:CoordinateSystem}	
\end{figure}

\subsubsection{Sweeps}
As depicted in \cref{fig:ExtendedPrimitives} most extended primitives can be ascribed to sweeps. Thereby a 2-dimensional sketch is 'swept' along a 3-dimensional path, forming a 3D object. $\mathbf{C}(\xi)$ denotes the point on the sweep path $\mathbf{C}$ at the local path coordinate $\xi$.
The following steps describe a point-in-membership test for a point of interest $\mathbf{P}$ and a sweep for the special case that (i) the local basis $\mathbf{A}(\xi)$ equals the basis of the sketch $\mathbf{B}(\xi)$ and (ii) the local basis system  $\mathbf{A}(\xi)$ follows the tangent of the path with a suitable description, e.g. the Frenet base. For the general case, see \cref{sec:rotlocbasis}.

\begin{itemize}
	\item The closest point $\mathbf{{C}}(\xi_{cp})$ on the sweep path with respect to the corresponding point of interest $\mathbf{P}$, is computed either analytically, or, if this is not possible, using Newton's method to find a root of the function
	\begin{equation}
			f(\xi) = \mathbf{\dot{C}}(\xi) \cdot  (\mathbf{P}- \mathbf{C}(\xi)) = 0.
			\label{eq:CriteriaNewtonClosestPoint}
	\end{equation}
$f(\xi)$ is the dot product between the tangent vector $\mathbf{\dot{C}}(\xi)$ and the vector pointing from a curve point $\mathbf{C}(\xi)$ to the point of interest $\mathbf{P}$. Provided that a suitable starting value is available Newton's method then iteratively delivers the corresponding coordinates $\xi^{j+1}_{cp}$ of ever closer points $\mathbf{{C}}(\xi^{j+1}_{cp})$:
	\begin{equation}
			\xi^{j+1}_{cp} = \xi^{j}_{cp} - \frac{f(\xi^{j}_{cp})}{\dot{f}(\xi^{j}_{cp})} = \xi^{j}_{cp} -\frac{(\mathbf{\dot{C}}(\xi^{j}_{cp}) \cdot  (\mathbf{P}- \mathbf{C}(\xi^{j}_{cp}))}{\mathbf{\ddot{C}}(\xi^{j}_{cp}) \cdot  (\mathbf{P}- \mathbf{C}(\xi^{j}_{cp})) + |\mathbf{\dot{C}}(\xi^{j}_{cp})|^2},
				\label{eq:inverseMappingSpline}
	\end{equation}
	where $\mathbf{\dot{C}}$ and $\mathbf{\ddot{C}}$ denote the first and second derivative of the sweep path $\mathbf{C}$. The minimal requirement for continuity of the sweep path is $C^1$.\\
	\textit{Remarks:} 
	\begin{enumerate}[label=(\roman*)]
		\item In our implementation the initial values are found by evaluating the distance to the points of an approximation polygon.		
		\item It is possible that~\cref {eq:CriteriaNewtonClosestPoint} has multiple solutions. Among the corresponding points on the curve $\mathbf{C}(\xi _i)$ the one with the smallest distance to the point $\mathbf{P}$ has to be chosen. 		
		\item There is the unlikely possibility that~\cref {eq:CriteriaNewtonClosestPoint} delivers multiple solutions $\mathbf{C}(\xi _i)$ with all the same distance to the point  $\mathbf{P}$. For sweeps it does not matter which of theses points is selected to be $\mathbf{{C}}(\xi_{cp})$. This, however, does not hold for lofts (for a more detailed explanation refer to~\cref{fig:CriteriaNewtonClosestPoint}).	
		\item Non-linear cases are possible, where $C^2$ can not be provided, i.e. the curvature is not continuous, or worse cannot even be evaluated (e.g. knot-multiplicity for splines). For that reason we suggest to compute the first derivative $\dot{f}$ (see~\cref {eq:inverseMappingSpline}) by finite differences.	
	\end{enumerate}
	
	\begin{figure}[H]\centering
	\includegraphics[width=14cm]{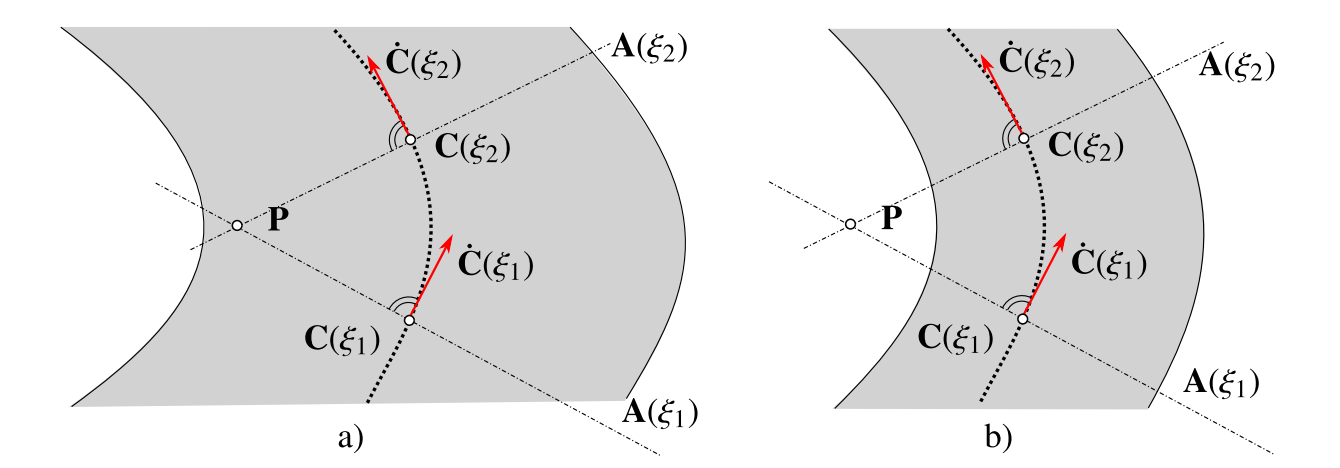}
	\caption{Multiple solutions for~\cref{eq:CriteriaNewtonClosestPoint} $\mathbf{C}(\xi_i)$ with the same distance to $\mathbf{P}$. For sweeps the sketch on plane $\mathbf{A}(\xi_1)$ coincides with the sketch on plane $\mathbf{A}(\xi_2)$. Hence, it does not matter on which plane the point-in-membership test is carried out. However, for lofts (see~\cref{sec:pimLoft}) the intermediate sketches are interpolated according to the arc length. In this case, the sketches at $\mathbf{A}(\xi_1)$ and $\mathbf{A}(\xi_2)$ will be different. Consequently all solutions for~\cref{eq:CriteriaNewtonClosestPoint} must be evaluated. The point is defined to lie inside if one sketch delivers this result. However, these cases occur only in the unusual cases, where the thickness of the body is large compared to the curvature (see case a)). More likely is case b) where all ambiguous solutions for~\cref{eq:CriteriaNewtonClosestPoint} lie outside.}
	\label{fig:CriteriaNewtonClosestPoint}	
\end{figure}
		
	\item The closest point $\mathbf{{C}}(\xi_{cp})$ forms the origin of the newly created local coordinate system $\mathbf{A}(\xi_{cp})$. To this end, the tangent vector of the curve is evaluated at $\mathbf{{C}}(\xi_{cp})$  and provides the z-axis of the local system $\mathbf{A}(\xi_{cp})$ using e.g. the Frenet base. Other alternatives are possible and presented in~\cref{sec:rotlocbasis}.
	
	\item The point of interest $\mathbf{P}$ is mapped into the local coordinate system $\mathbf{A}(\xi_{cp})$ to get $\mathbf{\tilde{P}}$ where $\tilde{P}_z = 0$ (see \cref{eq:PointMapping}).
	
	\item For the case that the local system $\mathbf{A}(\xi)$ coincides with $\mathbf{B}(\xi)$, a point-in-membership test can be performed with the mapped point $\mathbf{\tilde{P}}$ and the sketch contour line as the boundary. To this end, 2D ray casting is used.
	
\textit{Remark:} A considerable speedup, especially for spline contour lines, can be achieved if the sketch is modeled as quadtree of certain accuracy (similar to \cref{fig:octree}) for each extended primitive. This can be carried out once at the beginning of the analysis phase.
\end{itemize}

\begin{figure}[H]\centering
	\includegraphics[width=11cm]{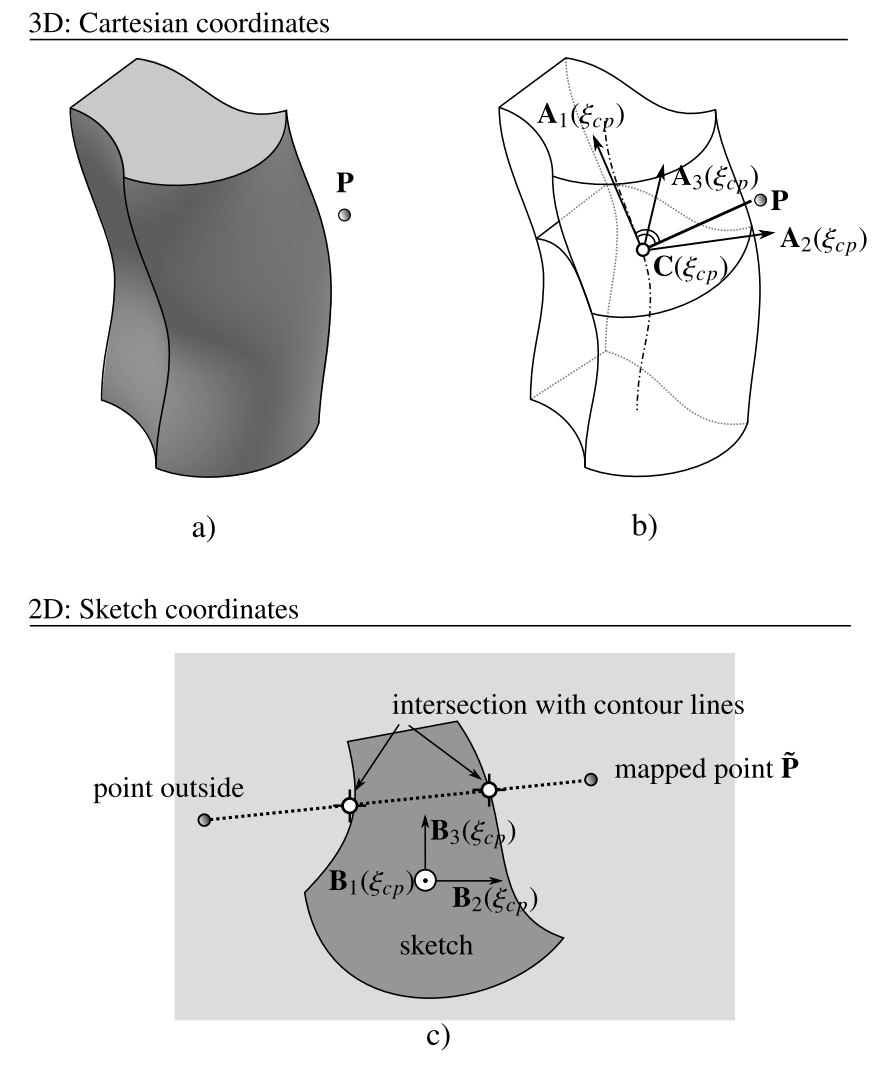}
	\caption{ Point-in-membership test of a sweep whose local basis $\mathbf{A}$ coincides with the sketch basis system $\mathbf{B}$: (a) Sweep with point of interest. (b) Form local basis $\mathbf{A}$ at closest point $\mathbf{C}(\xi_{cp})$ and (c) perform a point-in-membership test on the 2D sketch plane -- here, by ray casting with the contour line of the sketch.}
	\label{fig:PIM}	
\end{figure}

\subsubsection{Lofts} \label{sec:pimLoft}
\begin{figure}[H]\centering
	\includegraphics[width=14cm]{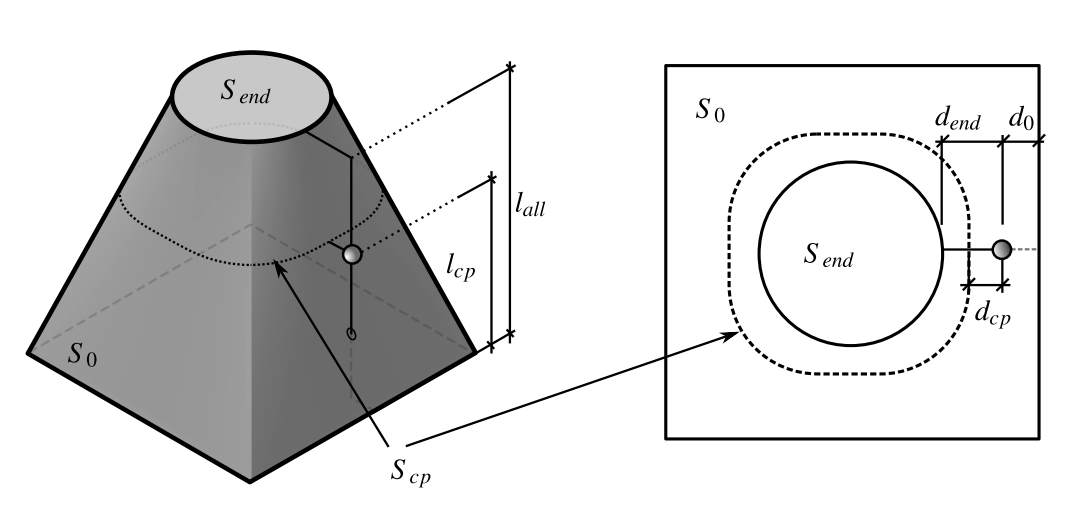}
	\caption{ Point-in-membership test on intermediate sketch $S_{cp}$ of a loft.}
	\label{fig:Loft}	
\end{figure}

Lofts can be treated very similarly to sweeps. Again, for simplicity, the loft path is considered to be orthogonal to the starting $S_0$ and ending $S_{end}$ sketch. For a loft, in contrast to sweeps, the 2D point-in-membership test must be performed on both the starting as well as the ending sketch. Additionally, the smallest distances to both contour lines $d_0$ and $d_{end}$ are calculated. If the mapped point $\mathbf{\tilde{P}}$ lies inside one sketch and outside the other, the distance to the intermediate contour line $d_{cp}$ is interpolated according to the arclength $l_{cp}$ of the closest point on the loft path (see \cref{fig:Loft}) and the overall length of the loft path $l_{all}$. In our implementation, linear interpolation is used. The distance to the point outside is set negative and, thereby, according to the sign of the interpolated distance $d_{cp}$, the point is outside for negative and inside for positive values. For the linear interpolation, the point-in-membership test reads
\begin{equation}
	\mathbf{P} \in B_{\text{Loft}} \quad \text{iff} \quad \quad d_i \geq 0
\end{equation}
with
\begin{equation}
	d_{cp} = d_0 + \frac{(d_{end}-d_0)}{l_{all}} l_{cp}.
\end{equation}

\subsection{Ray-casting with Splines (in 2D)} \label{sec:SplineIntersection}
\begin{figure}[H]
	\centering
	\includegraphics[width=14cm]{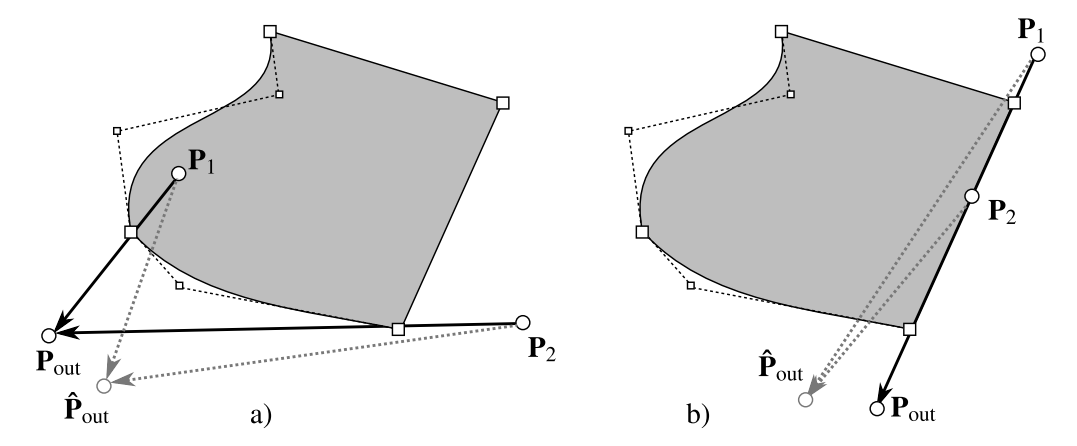}
	\caption{Problematic ray cast scenarios, where a) the ray intersects at a corner point and b) the ray is collinear with a contour line. However, a simple shift of the reference point ($ \mathbf{P}_{\mathrm{out}} \rightarrow \mathbf{\hat{P}}_{\mathrm{out}}$) can resolve these problems.}
	\label{fig:problemsRayCast}	
\end{figure}

As presented in~\cref{fig:PIM}, a point-in-membership test can be performed with ray-casting on a dimensionally reduced model, i.e. in 2D. This is by far simpler than carrying out ray-casting in 3D. On the two-dimensional intermediate sketch, a ray from a point that is definitely outside of the domain $\mathbf{P}_{\mathrm{out}}$ to the point of interest $\mathbf{P}$ is cast forth, and all intersections with the contour curves are counted. Possible elements for the contour lines are straight lines, arcs, or splines (B-Spline, NURBS). Problematic are the unlikely cases, in which the ray intersects the contour at a corner point (see~\cref{fig:problemsRayCast}(a)), or in which the ray is collinear with a contour line (see~\cref{fig:problemsRayCast}(b)) as the point membership is ambiguous. However, these cases can easily be detected and the algorithm simply changes the position of the reference point $\mathbf{P}_{\mathrm{out}}$.

\begin{figure}[h]\centering
	\includegraphics[width=16cm]{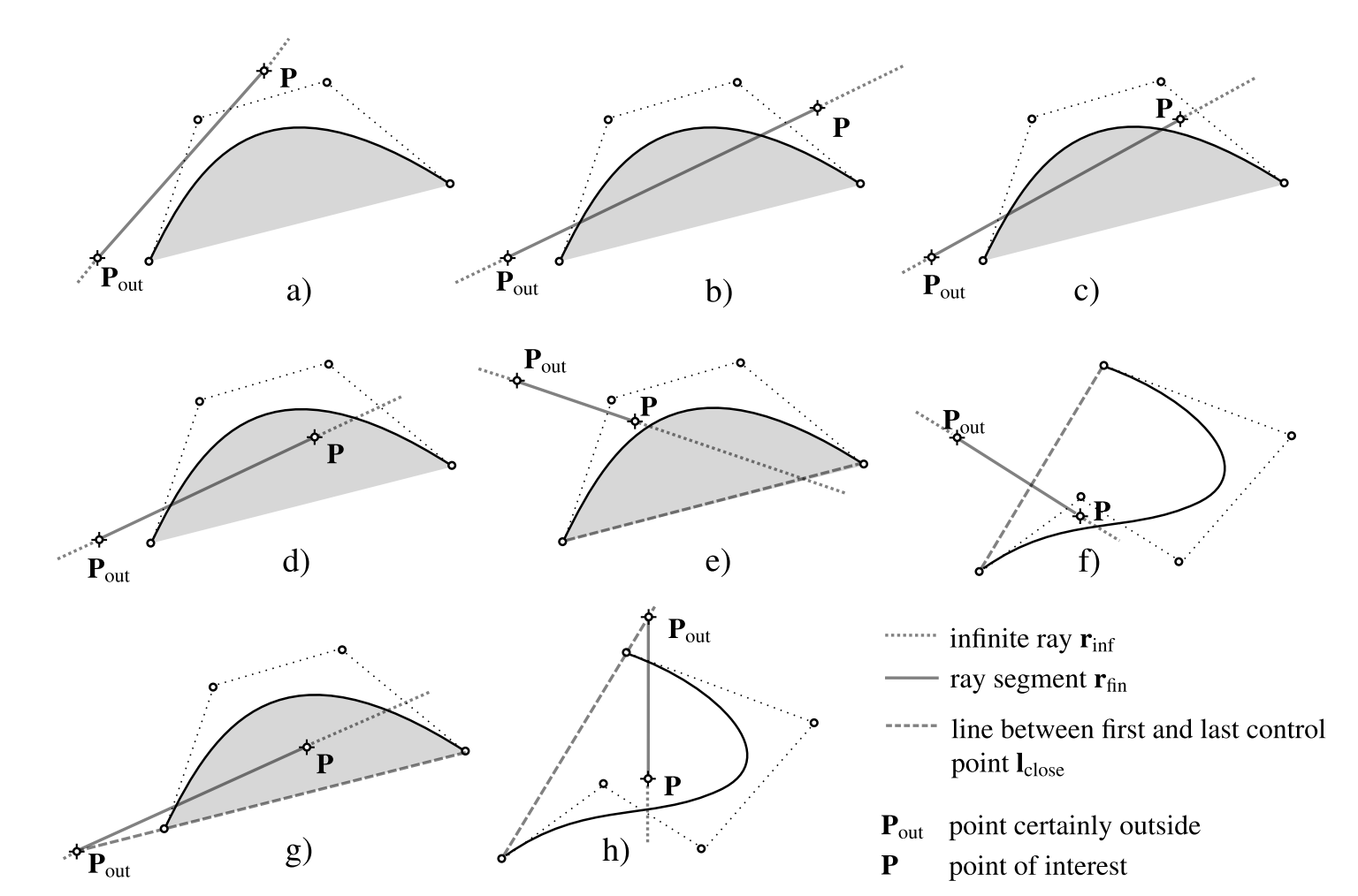}
	\caption{Different ray-cast scenarios: (a) Intersection only with control polygon. (b) Same number of intersections with spline as with control polygon. (c) Point between control polygon and spline (d) Point inside. In cases (e) and (f) a sole count of intersections with the control polygon will fail. To this end the infinite ray is also intersected with a line segment between first and last CP. At cases (g) and (h) the point outside $\mathbf{P}_{\text{out}}$ is set to be on the closing line $\mathbf{l}_{\mathrm{close}}$, thus, avoiding in most cases an expensive full intersection test. } 
	\label{fig:RayCastSpline}	
\end{figure}

 While there are analytical solutions to find the intersections of the ray and a line or an arc, the evaluation of the  number of intersections between a ray and a spline is not straightforward (see~\cref{fig:RayCastSpline}). These intersections can be found using a brute force method, which performs an intersection search with a fixed set of initial parameter values along the ray. This is very inefficient, as it is necessary to perform several closest point searches, containing several inverse mappings onto the spline, for each intersection search. Moreover, most of the initial values will find the same intersection points and, furthermore, it is not guaranteed that all intersections are found. Typically, Newton's method is used for the inverse mapping. Newton's method is highly dependent on the initial value. Hence, two intersections lying close to each other might be detected as only one intersection.  A robust and efficient way to obtain all zeros of B\'ezier curves with their multiplicity is presented in \citep{Machchhar2016}. However, the algorithm presented therein is specific to B\'ezier curves and not tuned to deliver the parity (i.e. whether the number of intersections is odd or even) of the intersections. The algorithm presented in this paper can use any spline description such as B\'ezier curves, B-Splines or NURBS. Moreover, as only the parity of intersections is needed, in most cases, an evaluation of the intersection points is not necessary. In this context, an efficient and robust method to find the parity of intersections of a (ray) line and a spline is depicted in \cref{fig:RayCastSpline2} and \cref{alg:rayCastSpline}.

As only the parity of intersections is needed, the convex hull property of the splines can be used. Let us consider cases a) and b) in \cref{fig:RayCastSpline}. In these cases, the ray segment (between the point of interest $\mathbf{P}$ and the point outside the model $\mathbf{P}_{\text{out}}$) has the same number of intersections as the infinite ray and, thus, determines the parity. This is due to the convex hull property which guarantees that the amount of intersection points between a line and a spline cannot be larger than the amount of intersections with a line and the spline's control polygon. There can be less intersections, but this does not change the parity of intersections. \\
The cases c) and d), as depicted in~\cref{fig:RayCastSpline}, are more sophisticated. Here, the ray segment $\mathbf{r}_{\mathrm{fin}}$ and an infinite ray $\mathbf{r}_{\mathrm{inf}}$ do not have the same number of intersections and, hence, it is necessary to determine intersection points of the ray with the spline. To this end, suitable starting values are needed for the Newton iteration. For the determination of these starting values several possibilities are available. One possibility is presented in \cref{sec:2dRayCast}. For simplicity of formulation, we assume that the ray corresponds to the positive x-axis. In other cases, a suitable transformation to a local coordinate system is performed (see \cref{sec:2dRayCast}). All zeros which fulfill the following property are intersection points with the spline, and their number determines the parity for the ray cast. \\
The cases e) and f), depicted in \cref{fig:RayCastSpline}, where the rays intersect with the line segment between the first and last control point (in the following denoted as $\mathbf{l}_{\mathrm{close}}$) are covered by an additional intersection test of infinite ray and line $\mathbf{l}_{\mathrm{close}}$. This, of course, leads to a different parity from the (finite) ray segment, and hence results in an expensive inverse mapping. Thus, it desirable to choose the point $\mathbf{P}_{out}$ in such a way, that the cases e) and f) are seldom. The best choice is to place the point $\mathbf{P}_{out}$ on to the line $\mathbf{l}_{\mathrm{close}}$ (see~\cref{fig:RayCastSpline}(g) and (h)). This might however not always be possible. In this context consider a contour, which is constructed of more than one spline. Even for the simple case of two splines, where most likely an intersection point of both closing lines exist, i.e. $\mathbf{l}_{\mathrm{close}_1}\cap \mathbf{l}_{\mathrm{close}_2} \neq \emptyset$ it is not guaranteed that this point also lies outside of the domain. In these cases the point outside will be set on $\mathbf{l}_{\mathrm{close}}$ of one spline, hence avoiding the expensive point-in-membership test at least for one spline.

%% file: numericalExamples.tex
\section{Numerical examples} \label{sec:numExp}
In this chapter, we present three different examples, which are created with extended primitives and extended operations. They are designed to provide an overview of the possibilities of the modeling approach presented in this paper.

\subsection{Sweep example} 
\begin{figure}[h]
	\centering
	\includegraphics[width=16cm]{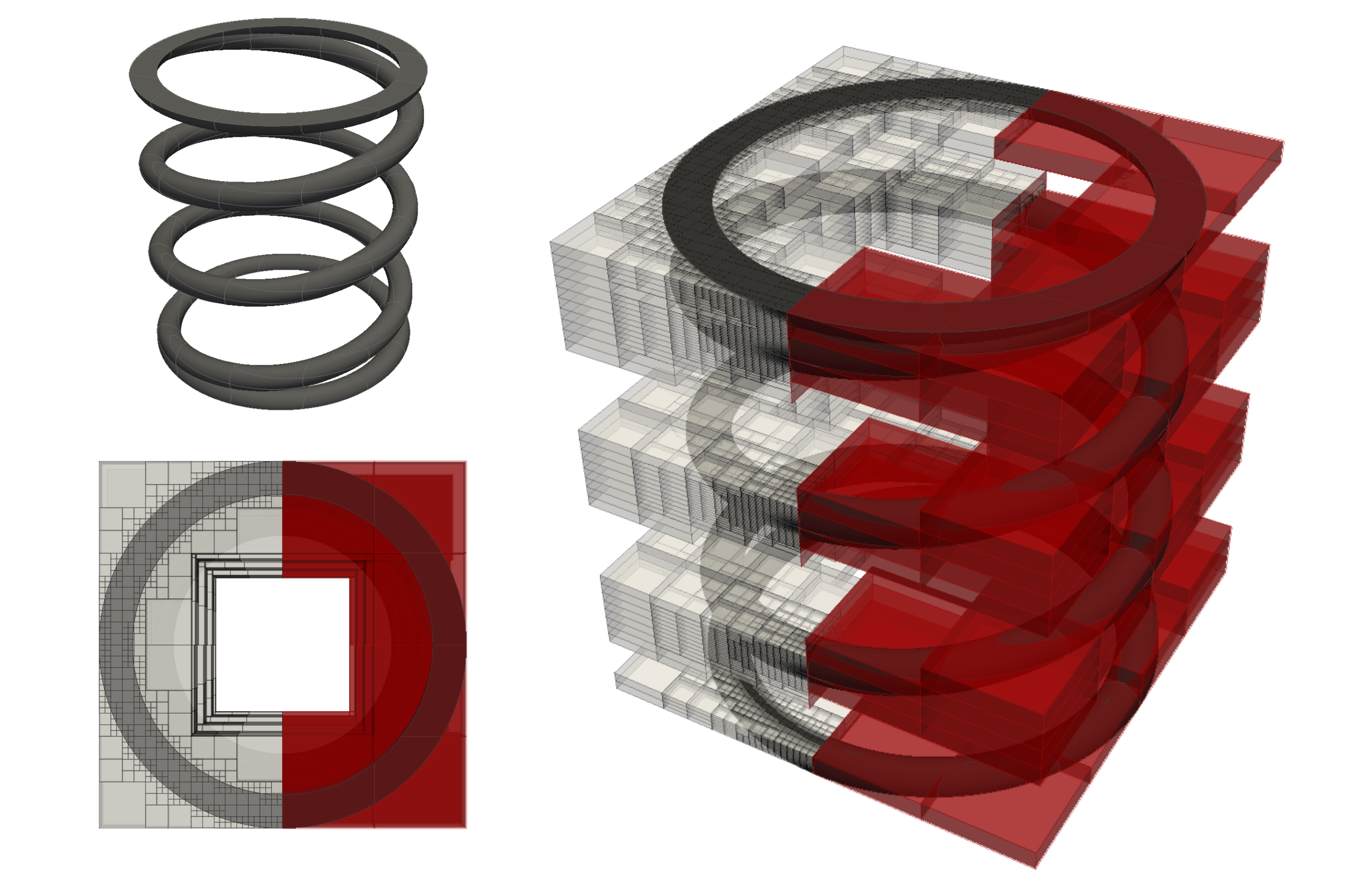}
	\caption{Coil spring model: Depicted are the finite cells (red) on the one side and the octree partitioning with a refinement depth of $k=4$ on the other side (light gray), respectively, for top and isometric view. }
	\label{fig:Spring}
\end{figure}

The first example involves a coil spring, which is constructed using sweeps (see \cref{fig:Spring}). The spring is constructed by three primitives, which are combined with two union operations. For all three bodies, a circle sketch with a radius $r_{\mathrm{sketch}}=1$ is swept along the corresponding sweep path which, for the bottom and top torus, is again a circle of radius $r_{\mathrm{torus}}=10$. Bottom and top tori are aligned to the $x-y$ plane and located at $z_{\mathrm{bottom}} = 0 $ and $z_{\mathrm{top}} = 24$, respectively. The computational domain ranges from $z=0$ to $z=1$, thus clipping bottom and top tori into half-tori. The sweep path of the coil is described by a helical NURBS of degree $p=2$, the knot vector
\begin{equation*}
	U = [0,0,0,1,1,2,2,3,3,4,4,5,5,6,6,7,7,8,8,9,9,10,10,11,11,12,12,12] \quad ,
\end{equation*}
and the 3D control points with weights $w_1 = 1$ or $w_2 = \frac{1}{\sqrt{2}}$
\begin{equation*}
	\mathbf{P}_i = \begin{pmatrix}
	x_i \\ y_i \\ z_i \\ w_i
\end{pmatrix}	= \left[ \begin{matrix}
		10 & 10 &  0 & -10 & -10 & -10 &   0 &  10   \\
		0  & 10 & 10 &  10 &   0 & -10 & -10 & -10   \\
		0  &  1 &  2 &   3 &   4 &   5 &   6 &   7   \\
		w_1& w_2& w_1& w_2 & w_1 & w_2 & w_1 & w_2
\end{matrix} \right. \quad ...
\end{equation*}
\begin{equation*}
  ... \quad \begin{matrix}
		10 & 10 &  0 & -10 & -10 & -10 &   0 &  10   \\
		0  & 10 & 10 &  10 &   0 & -10 & -10 & -10   \\
		8  &  9 & 10 &  11 &  12 &  13 &  14 &  15   \\
		w_1& w_2& w_1& w_2 & w_1 & w_2 & w_1 & w_2
\end{matrix} \quad ...
\end{equation*}
\begin{equation*}
  ... \left. \quad \begin{matrix}
		10 & 10 &  0 & -10 & -10 & -10 &   0 &  10 & 10 \\
		0  & 10 & 10 &  10 &   0 & -10 & -10 & -10 &  0 \\
		16 & 17 & 18 &  19 &  20 &  21 &  22 &  23 & 24 \\
		w_1& w_2& w_1& w_2 & w_1 & w_2 & w_1 & w_2 & w_1
\end{matrix} \right] \quad .
\end{equation*}

 For the simulation, we choose 4x4x24 finite cells with integrated Legendre shape functions and a polynomial degree of $p=7$ in the trunc space ~\citep{Szabo2004}. Cells that are located completely outside the coil are deactivated, reducing the number of active finite cells from 384 to 134. The integration of the element matrices is carried out using an adaptive octree, whose maximum partitioning depth is preset to $k=6$.

\begin{figure}[H]\centering
	\includegraphics[width=16cm]{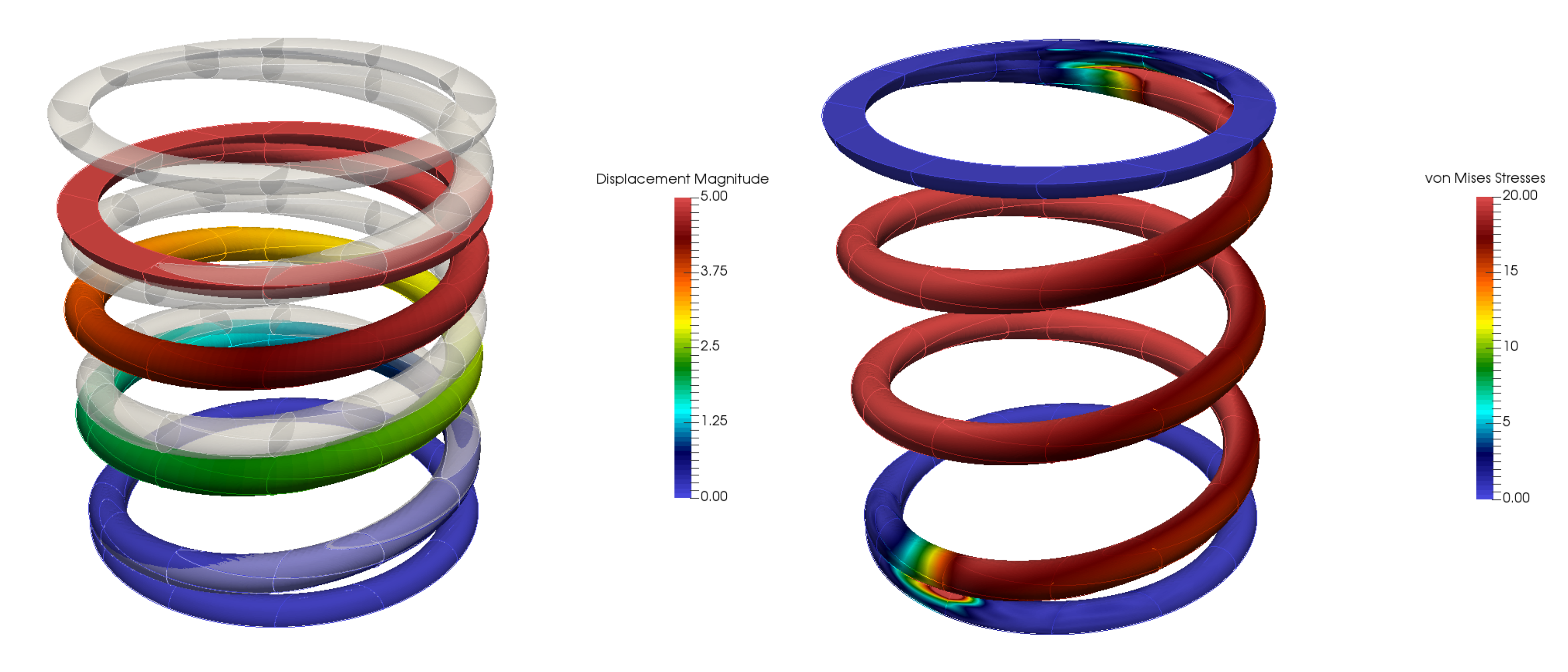}
	\caption{ Displacements and von Mises stresses of the coil spring under predefined top displacement of $\mathbf{\hat{u}}_z = -5$. }
	\label{fig:CoilResults}	
\end{figure}

In this model, strong Dirichlet boundary conditions are applied at the top and bottom faces of the model, which correspond to the boundaries of the finite cells. The degrees of freedom here are fixed in all directions, except for the vertical displacement on the top face which is predefined by $\mathbf{\hat{u}}_z = -5$. \Cref{fig:CoilResults} shows the resulting displacement and the von Mises stresses, which provide a good overall insight into the structural load carrying behavior. They are smooth throughout most part of the domain and exhibit singular behavior at the intersection curves of  the top and bottom tori with the coil. As in any finite element computation, an accurate solution of local stresses at singularities can not be accurately resolved without an appropriate refinement. A local refinement is not carried out for this example, but it is possible, e.g. by application of the multi-level $hp$-method recently proposed in~\citep{Zander2015}.

It is noteworthy that only the CSG model is used in all involved steps, i.e. from the set-up of the model until the computation itself. A conversion from the CSG-model to an explicit B-rep is only carried out for the visualization of the results in the post-processing. Here, the marching cubes algorithm~\citep{Lorensen1987} is used to derive a triangulated surface on which the results are post-processed. However, even this conversion for visualization is not mandatory as volumetric post-processing is a possible option as well. 

\subsection{Loft example}
\begin{figure}[H]\centering
	\includegraphics[width=16 cm]{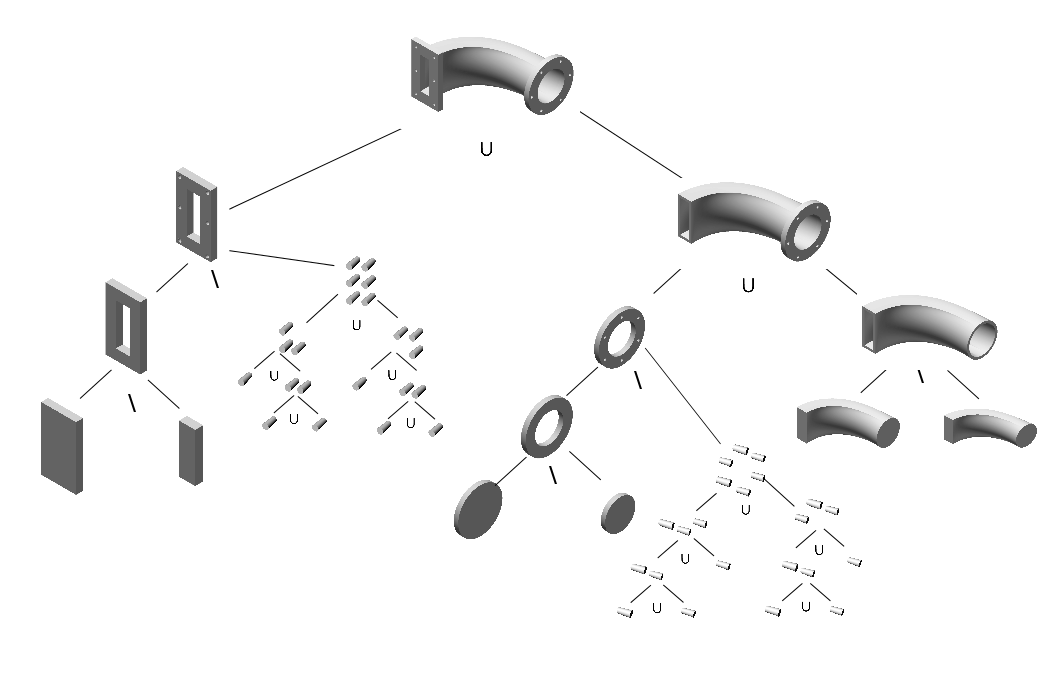}
	\caption{ CSG tree of loft.}
	\label{fig:LoftCSGTree}	
\end{figure}

\begin{figure}[H]\centering
	\includegraphics[width=12 cm]{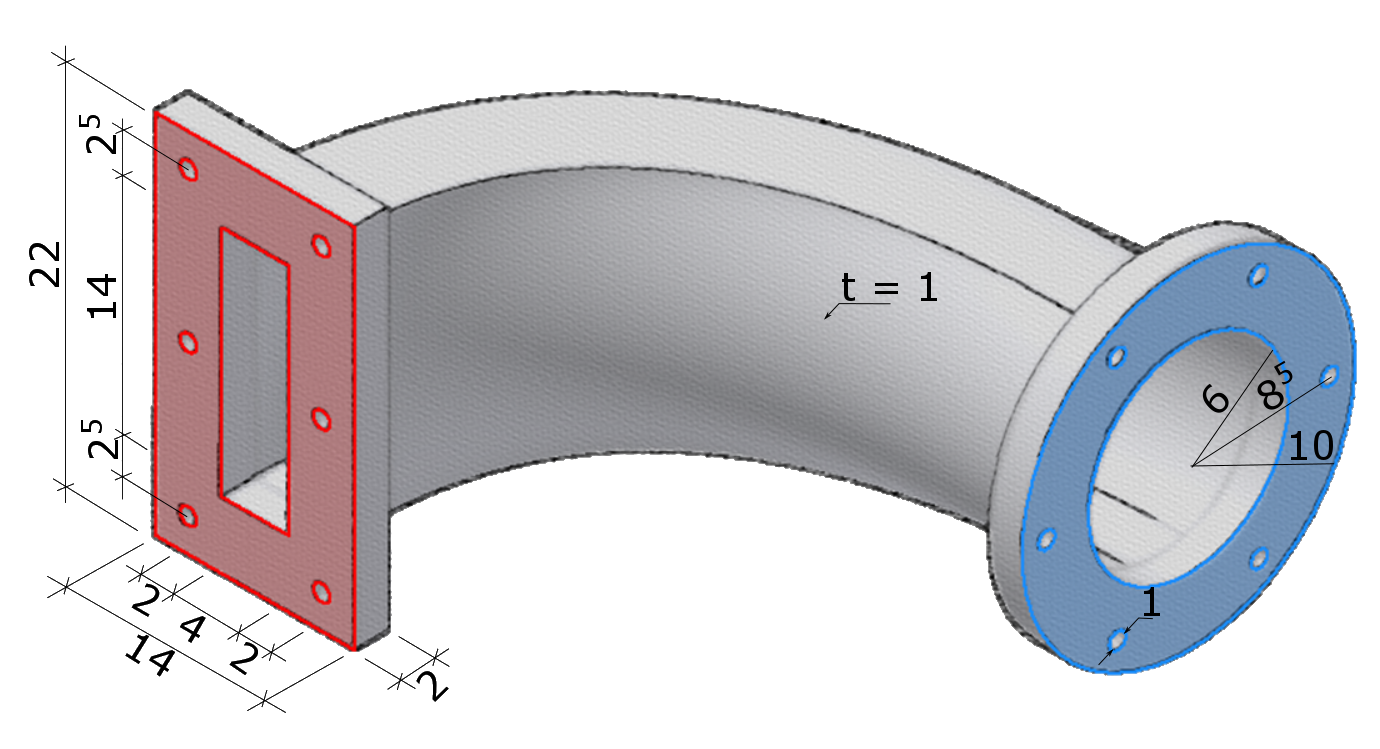}
	\caption{Dimensions of the loft.}
	\label{fig:DimesionsLoft}	
\end{figure}

The second example is a pipe elbow starting with a circular profile and ending with a rectangular cross section. It is constructed as a procedural model and then transformed to a CSG tree (see \cref{fig:LoftCSGTree}). It combines several simple primitives (cylinders and cuboids) and two lofts along a quadratic B-Spline loft path. The dimensions of the example are depicted in \cref{fig:DimesionsLoft}.

\begin{figure}[H]\centering
	\includegraphics[width=12cm]{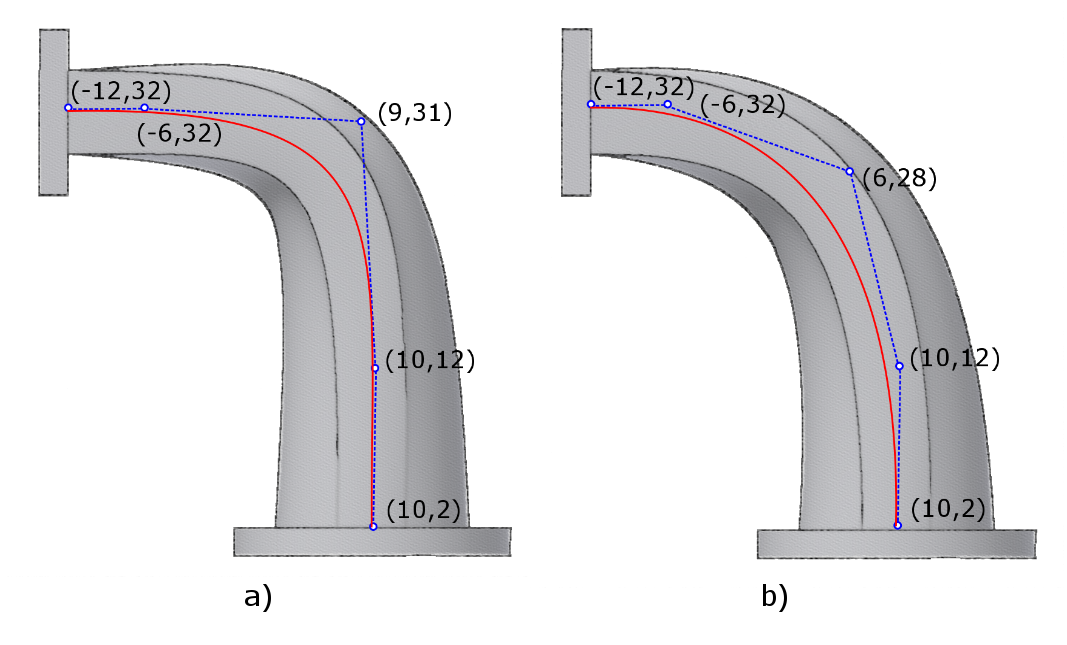}
	\caption{(a) Loft path of the first model. (b) Loft path of the second model.}
	\label{fig:MesureSide}	
\end{figure}

Again, two models are created to show another advantage of the FCM and its capability to use the explicit volume description of CSG models. In parametric design, even a slight change of few parameters may result in a re-meshing of parts, or even the entire model. In this example, the position of one control point of the loft path is changed. This results in a slightly different model (see  \cref{fig:MesureSide}). As this only leads to a change in the geometry, not in the topology, the mesh for a FEM-simulation does not necessarily require a re-meshing -- but it could be morphed. However, in the proposed work-flow, the change in the geometry does not need any special treatment: As it is only the loft path that changes, the same CSG-tree can be used with the same computational mesh. The change of the geometry is resolved on the integration level, leading to a different composed integration.

The resolution of the model is chosen to be 20x15x20 cells with integrated Legendre shape functions and a polynomial degree of $p=5$. As in the previous example, it is sufficient to consider cells that contain parts of the physical domain. The base plate (round, blue) is fixed, and a predefined deflection $\mathbf{\hat{u}} = 1$ is applied onto the left base plate (rectangular, red) using strong Dirichlet boundary conditions. For a precise integration of the stiffness matrix, the cells are partitioned with an octree to a maximum depth of five subdivisions. 

\begin{figure}[H]\centering
	\includegraphics[width=16cm]{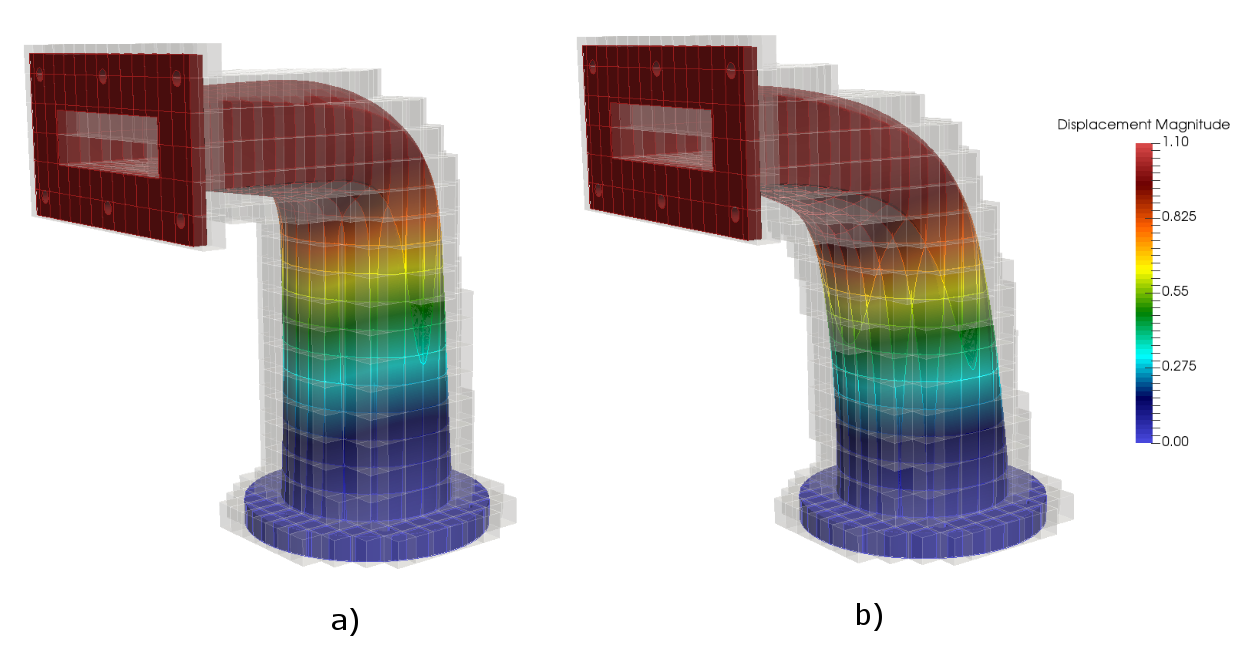}
	\caption{ Displacements for the (a) original example and the (b) modified example. }
	\label{fig:LoftDispl}	
\end{figure}

\begin{figure}[H]\centering
	\includegraphics[width=16cm]{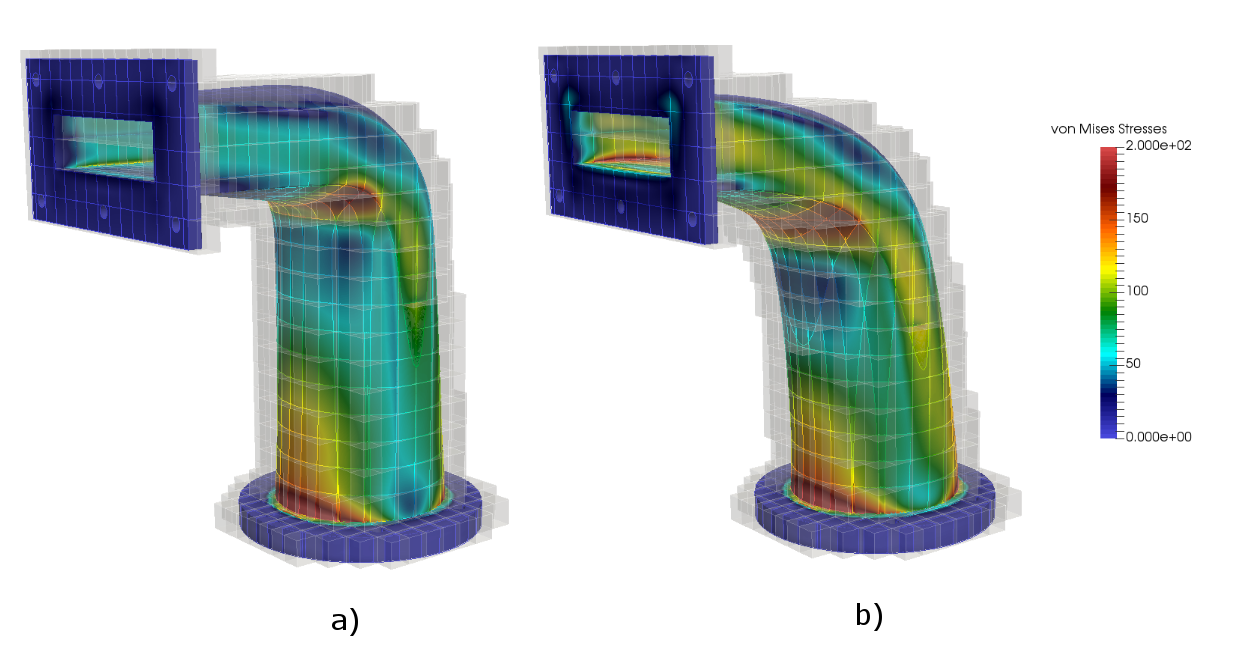}
	\caption{ Von Mises stresses for the (a) original example and the (b) modified example. }
	\label{fig:LoftStress}	
\end{figure}

\Cref{fig:LoftDispl} shows the finite cells embedding the structural model and the computed displacements for both examples. The von Mises stresses are depicted in~\cref{fig:LoftStress}.

\begin{figure}[H]\centering
	\includegraphics[width=12cm]{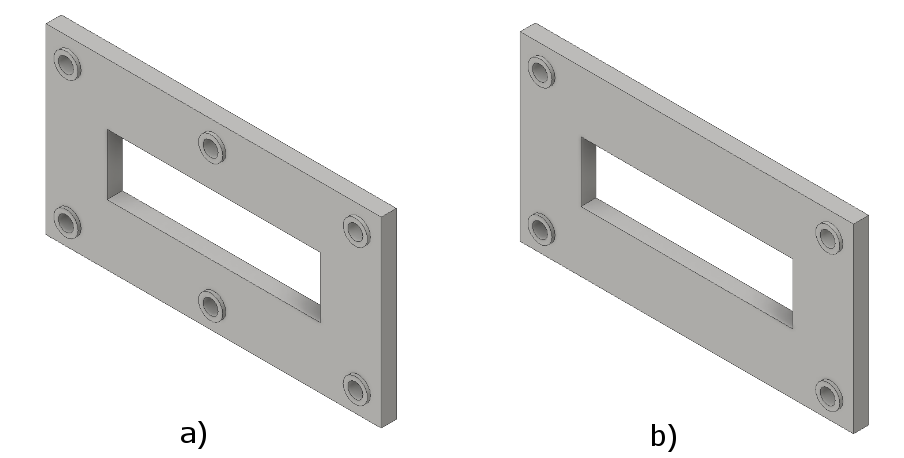}
	\caption{ Different topological models: (a) Rectangular plate with 6 holes (b) and 4 holes. }
	\label{fig:plateWithHoles}	
\end{figure}

As pointed out before, a geometrical change does not influence the proposed work-flow. In this case, even a topological change has hardly any impact. In a classical FEM-simulation, however, a re-meshing of the affected region would be required. A simple modification of the model is applied to illustrate a topological change. The number of holes in the rectangular plate is set to (i) six in the first and (ii) four in the second case. A Neumann boundary condition is applied to the holes –- via added washers at each hole. The loaded surfaces of the washers are automatically recovered using the marching cubes algorithm. A pressure of $\hat{p}=1$ is deployed to the upper row of washers, and a pressure of $\hat{p}=-1$ to the lower row, inducing a moment onto the rectangular plate (see \cref{fig:plateWithHoles}).

\begin{figure}[H]\centering
	\includegraphics[width=14cm]{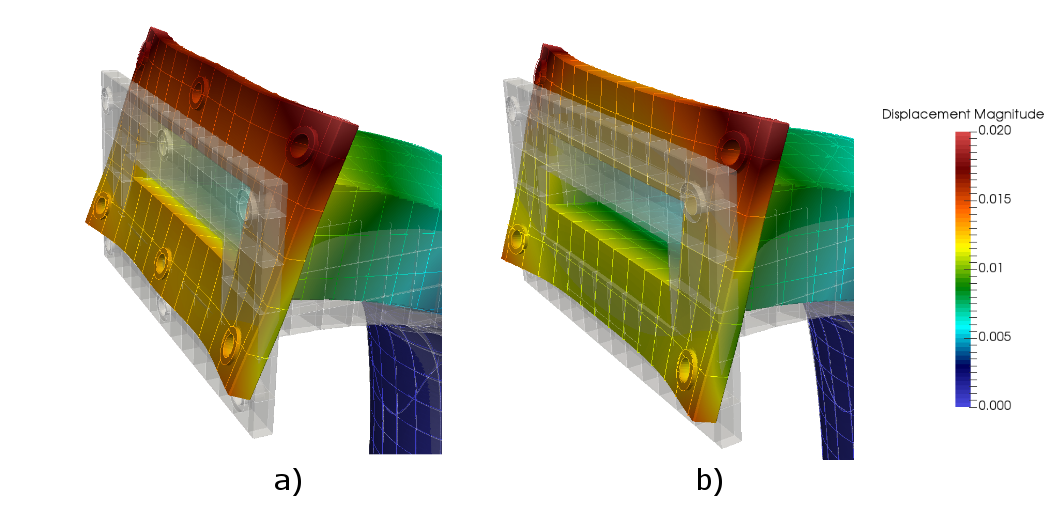}
	\caption{  Displacements at the rectangular plate for the (a) first example and the (b) second example. (Displacements scaled by a factor 200) }
	\label{fig:washerDispl}	
\end{figure}

\begin{figure}[H]\centering
	\includegraphics[width=14cm]{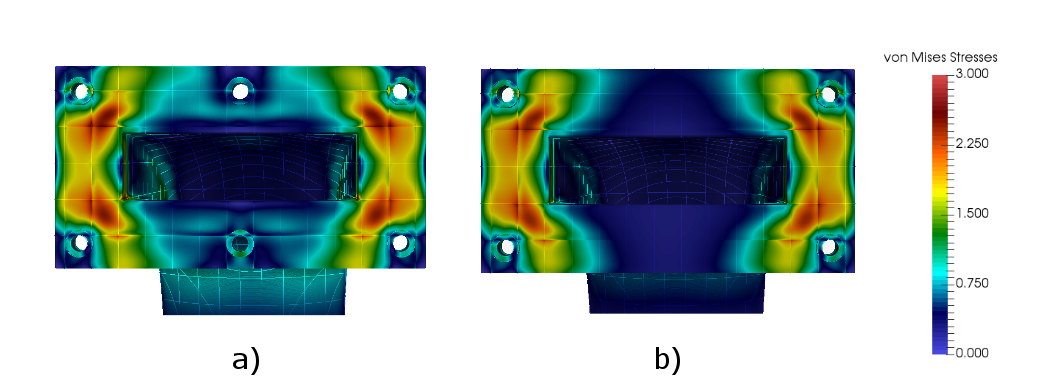}
	\caption{  Von Mises stresses at the rectangular plate for the (a) first example and the (b) second example. }
	\label{fig:washerStress}	
\end{figure}

\Cref{fig:washerDispl} shows the displacements and \cref{fig:washerStress} von Mises stresses at the rectangular plate for the two topologically different examples.

\subsection{Extended operations example}

\begin{figure}[h]\centering
	\includegraphics[width=14cm]{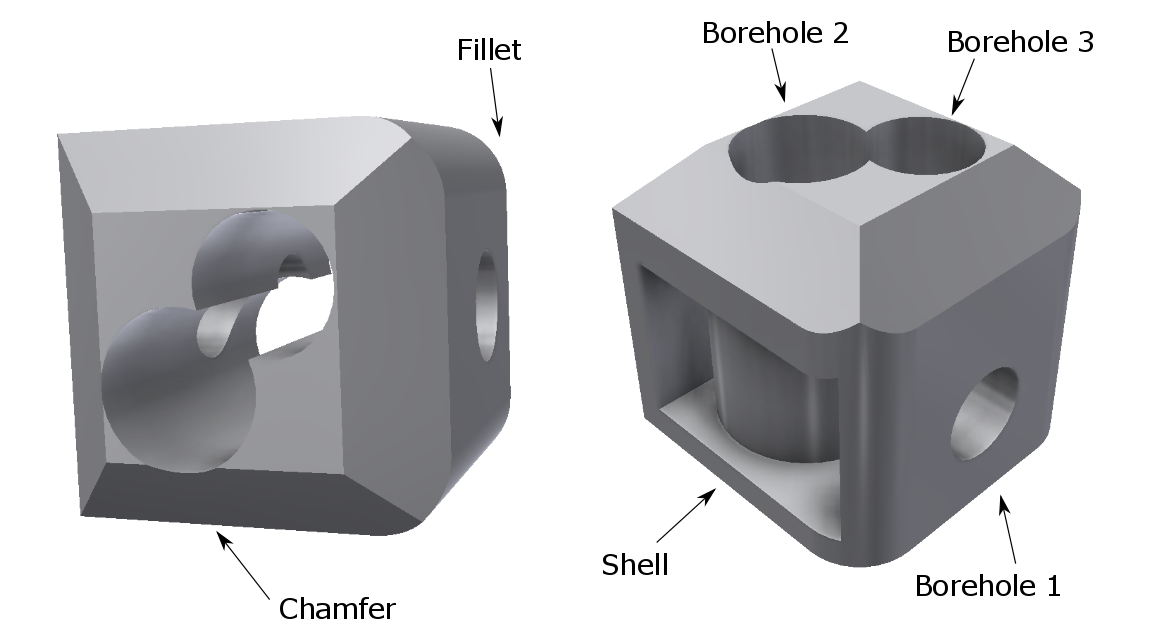}
	\caption{ Extended operations applied on a cube: chamfer, fillet, drilling a hole and shell operation. }
	\label{fig:OperationsModel}	
\end{figure}

The third example deals with commonly used extended operations. Starting from a cube, four edges are chamfered, two edges are filleted, three holes are drilled, and one shell operation is applied (see~\cref{fig:OperationsModel}).

\begin{figure}[H]\centering
	\includegraphics[width=6cm]{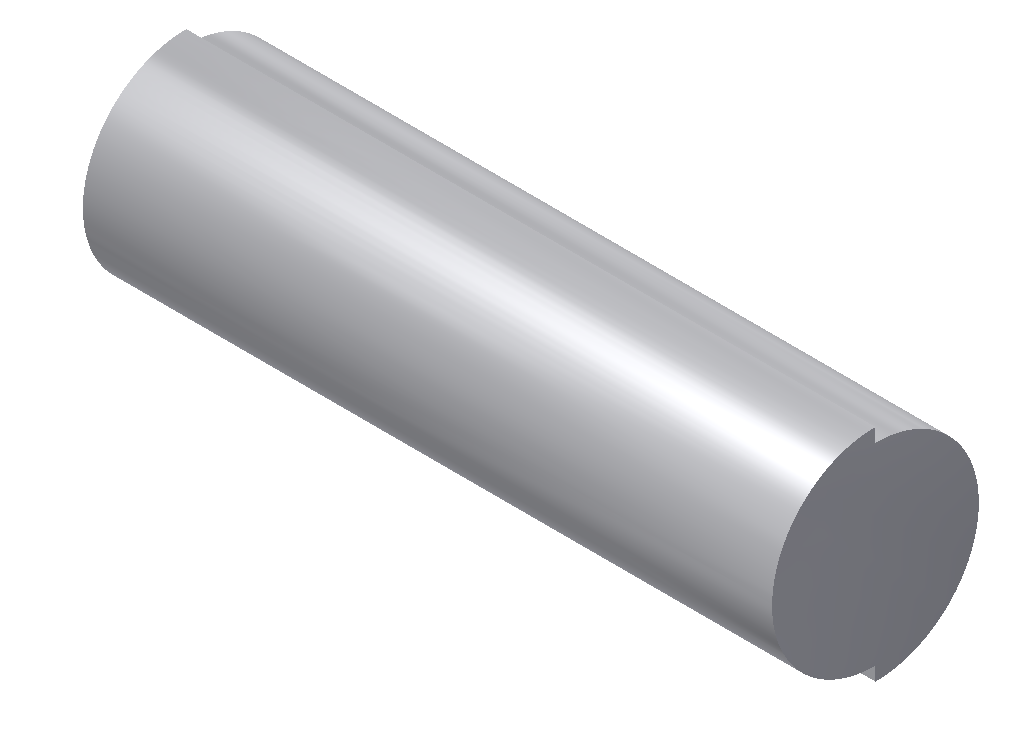}
	\caption{ For drilling the borehole 1 in the second model: Two slightly shifted cylinders (Shift here not to scale!).}
	\label{fig:HalfCylinders}	
\end{figure}

\begin{figure}[H]\centering
	\includegraphics[width=16cm]{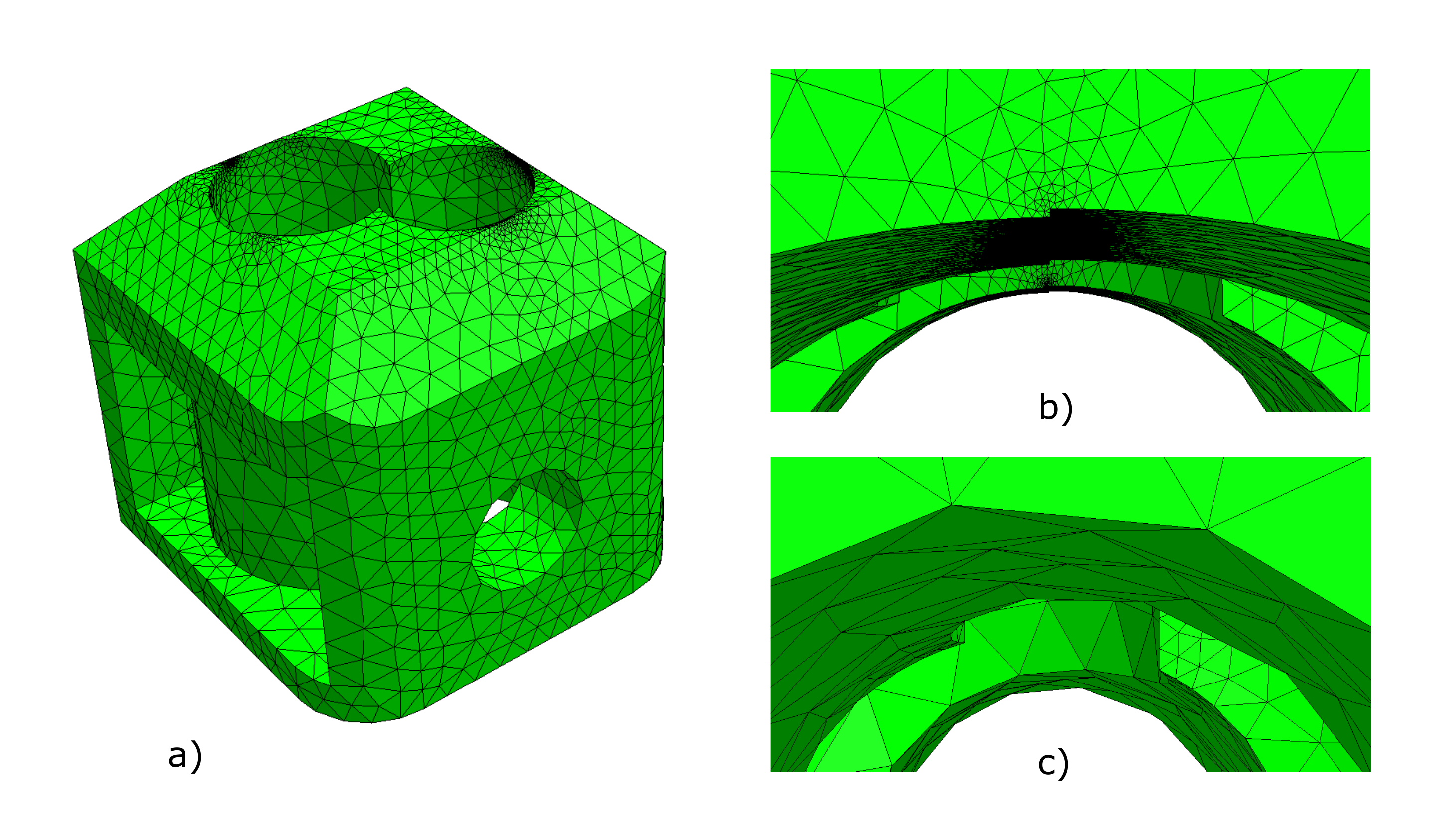}
	\caption{ Model meshed using Netgen ~\cite{Schoberl2003}: (a) Global view of the mesh, (b) refinement around the hole, which is constructed by two slightly shifted half-cylinders and (c) refinement around a hole without the (unnecessary) geometric detail.}
	\label{fig:Meshes}	
\end{figure}

As mentioned in~\cref{sec:CSGModelling}, the extended operators fillet, chamfer, and drilling a hole are just a combination of the original three boolean operations. Shells, however, are not a straightforward extension. The shell operation is applied to one surface, caving the volume while keeping a defined wall thickness to all other surfaces. In the present model, it is possible to manually model the shell operation with a set of Boolean operations and additional primitives by user interaction. This, however, is not straightforward for more sophisticated cases. \\
 Two almost identical models serve to show the capability of the FCM. In the first model, the (solitary) borehole 1 is created by subtracting a cylinder. In the second model, the hole is created by two extruded half cylinders, which are shifted by 0.05\% w.r.t. to the extension of the entire model (see \cref{fig:HalfCylinders}). This shift does not have any significant influence on the results of the simulation, but it does lead to a considerable increase in the effort needed for the classical FEM, where body fitted meshes must be used. Figure \ref{fig:Meshes} shows a mesh created by Netgen \citep{Schoberl2003} and the different refinements in the region around the holes. Although the shift is very small and only applied at one hole, the mesh of the second model has around 18 times more elements. Moreover, many elements are very badly conditioned. Even if this mesh represents the 'exact' geometry of the CAD model, it is very probable that it is not the intended finite element discretization. Similarly, there are often seemingly unmotivated refinements in practical applications, which is why considerable engineering efforts have to be made to remodel a structure before an efficient numerical analysis can be carried out.
 
\begin{figure}[H]\centering
	\includegraphics[width=16cm]{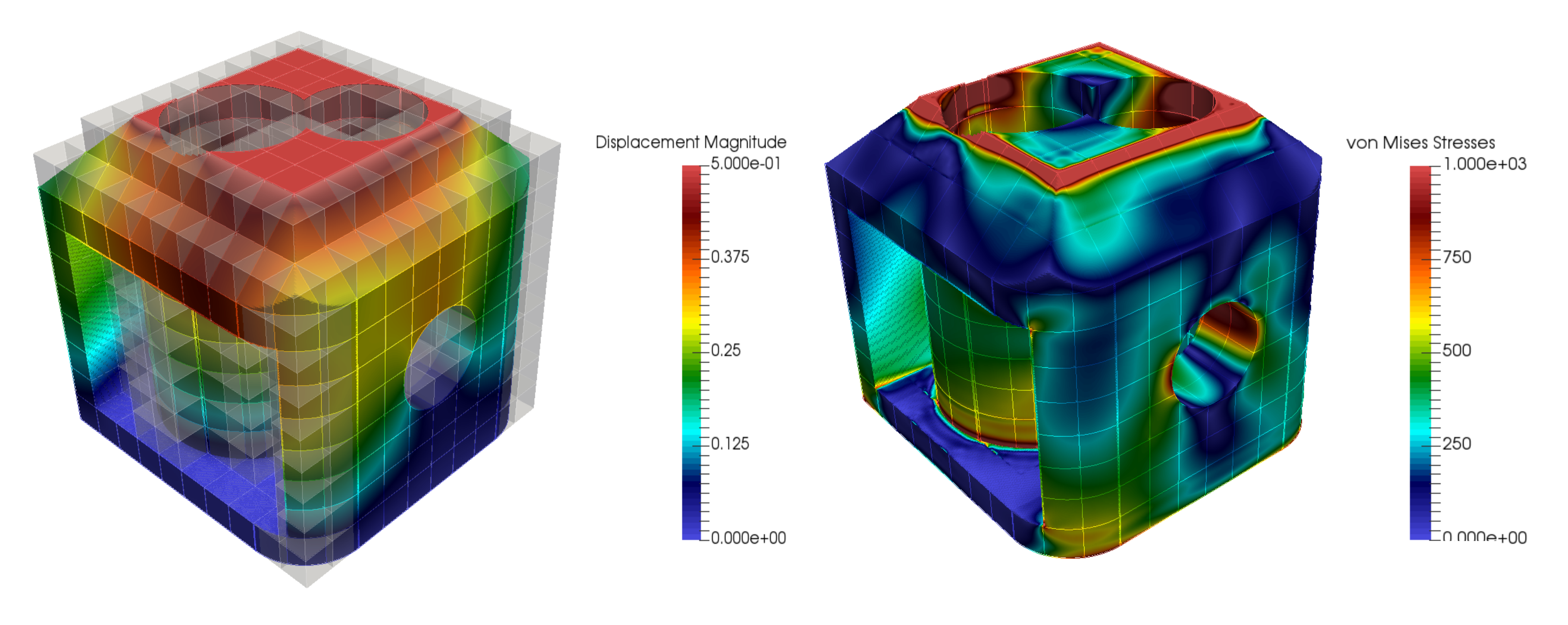}
	\caption{ Displacements with active cells and von Mises stresses }
	\label{fig:CubeResults}	
\end{figure}

\begin{figure}[H]\centering
	\includegraphics[width=12cm]{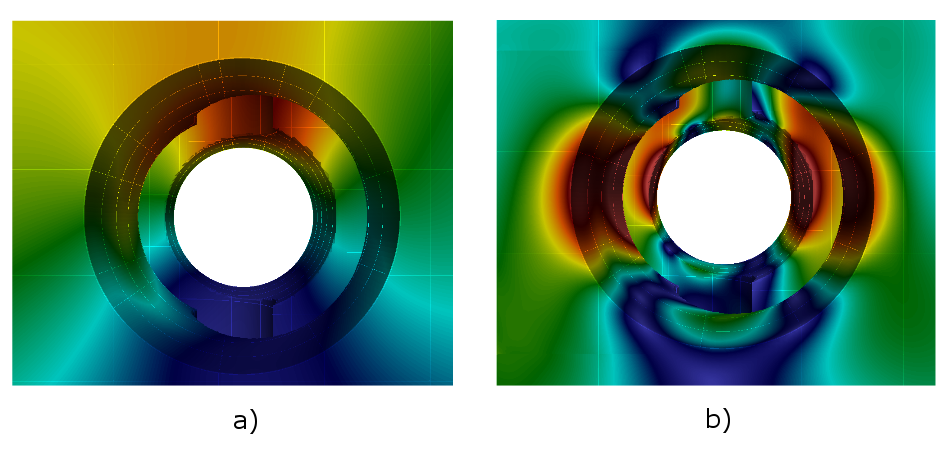}
	\caption{ Close up at borehole 1: (a) Displacements and (b) von Mises stresses }
	\label{fig:CloseUp}	
\end{figure}

For the FCM simulation, 10x10x10 finite cells using integrated Legendre polynomials of degree $p=5$ and a octree partitioning depth of $k=4$ are used. Again, the lower face is fixed, and the top face is displaced downwards by a value of $dz = -0.5$ using strong Dirichlet boundary conditions. Figure \ref{fig:CubeResults} shows the displacements and von Mises stresses of the model. A close-up of the displacements and von Mises stresses around the shifted hole is depicted in \cref{fig:CloseUp}. The results of the FCM computation for the two models (full cylinder versus two shifted half-cylinders to create borehole 1) are identical up to machine precision ($10^{-9}$), proving that the proposed method is robust against imprecise CAD modeling. 

%% file: conclusions.tex
\section{Conclusions} \label{sec:conclusions}
For the industry, integrated design processes and numerical analyses without complex transitions such as meshing are of high relevance, and many research groups have focused on this topic during the last years. While Isogeometric Analysis provides an excellent method for the numerical simulation of boundary representation models and shell-like structures, this paper focused on models created with Constructive Solid Geometry. A method has been presented that leads straight from CAD modeling to a numerical simulation using CSG and the Finite Cell Method (FCM). FCM is able to use the implicit model description provided by the CSG model directly, which simpliﬁes the meshing process significantly. It was shown that point-in-membership tests can be carried out efficiently for extended primitives like sweeps, bodies of revolution, and lofts. Also, extended operations such as fillet, chamfer, and holes can be applied to the model easily. Using FCM as a simulation technique that can deal with the explicit geometry description of CSG models has several advantageous properties. Following the design-through-analysis idea, a meshing step can be skipped. Also, as the CSG is inherently watertight, the geometry does not have to be pre-processed. Furthermore, CSG provides information about the interior of the models.\\
Although CSG modeling does not support the modeling of free form surfaces and objects, it is still possible to include these as B-Rep primitives to the CSG tree, provided that they are watertight and thereby support a reliable point membership classification.\\
Nevertheless, the presented methodology has also some drawbacks. Shell operations require user action and are typically not straightforward. Another issue is the conversion from a sequential model to a CSG tree. In our implementation, the CSG tree is constructed parallel to the information input of the sequential model, resulting in a sub-optimal tree, which is typically of high depth. However, it is possible to rearrange the CSG-tree in a way that allows to skip many point-in-membership tests. Further challenges are the complex curve descriptions for sweeps and lofts by B-Splines and NURBS. As no explicit inverse mapping is available, the point-in-membership test for bodies defined by these curves can become computationally  costly. Several approaches were made to improve the efficiency, among others the improved intersection test on B-Splines and NURBS (see Chapter \ref{sec:SplineIntersection}), which can avoid most of the inverse mappings. Also, the adoption of intersections of primitives with their bounding boxes led to a considerable speedup.

%% file: appendix.tex
\appendix
\section{Appendix}

\subsection{Rotated local basis system } \label{sec:rotlocbasis}
\begin{figure}[H]\centering
	\includegraphics[width=16cm]{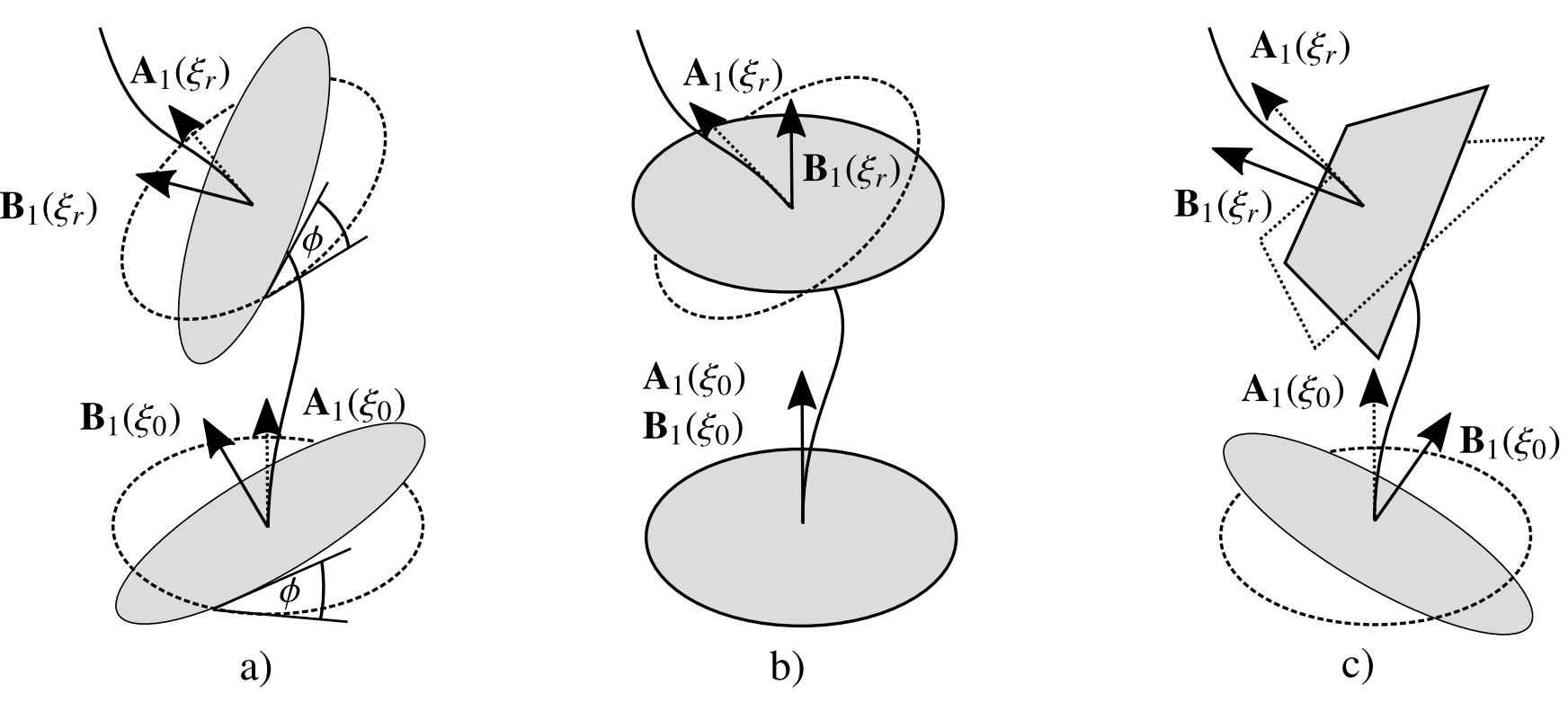}
	\caption{ Considering a sweep or loft: (a) Sketch rotated local basis system of the sweep path under a certain constant dihedral angle $\phi$. (b) Sketches always parallel to starting sketch. (c) Local coordinates system only known for starting and ending sketch.}	
	\label{fig:SweepSoph}	
\end{figure}

In contrast to the cases presented in \cref{sec:PIM_Extended} the point-in-membership test is more sophisticated for the following cases (see \cref{fig:SweepSoph}). Case (a) shows a sweep or loft whose sketch plane is not orthogonal to the sweep path at the starting point  $\mathbf{{C}}(\xi_{0})$. Nevertheless, the local basis of the sketch $\mathbf{B}$ and the basis spanned by the sweep path $\mathbf{A}$ are fixed in a certain relation, i.e. the dihedral angle $\phi$ between $\mathbf{A}$ and $\mathbf{B}$ remains constant along the path. In case (b), the normal of the intermediate sketches does not follow the tangent of the sweep path, but both starting and ending sketch remain parallel to each other. Finally, in case (c), the normal orientation of the sketch is only known at the starting and the ending sketch, and it changes along the sweep path. This latter case is most likely to occur during the construction of lofts where the normal of the starting and ending sketch often do not have the same angle to the tangent vector. 

For these three cases, an intermediate plane cannot be created at the closest point, but must be created in such a way that the $z$-coordinate of the mapped point $\mathbf{\tilde{P}}$ is zero. The local sketch basis spanned here is denoted by $\mathbf{B}({\xi_r})$.
\begin{equation}
	\mathbf{\tilde{P}} = 
	\begin{bmatrix}
		\tilde{P}_x \\
		\tilde{P}_y \\
		\tilde{P}_z \overset{!}{=} 0
	\end{bmatrix}.
\end{equation}

In the following, the steps to determine the intermediate sketch basis $\mathbf{B}(\xi_r)$ for the three cases are presented.

\begin{algorithm}
	\caption{Case (a): Rotated local basis systems under constant dihedral angle}	 \label{alg:fixedDihedral}
	\begin{algorithmic}[1]
			\State Let $\xi_0$ be the parameter value of the starting point of the sweep path
			\Statex
			\State \textit{0. Express at $\xi_0$ the base vectors of the sketch $\mathbf{B}_i(\xi_0)$ in terms of the local base of the path $\mathbf{A}_i(\xi_0)$.}
			\State \textit{Comment: This needs only to be done once at the set-up of the sketch}
			\Procedure{Find $\mathbf{A}$-$\mathbf{B}$ relation}{}
				\State	Compute 3x3 transformation matrix $\mathbf{T} =		\begin{bmatrix}	\mathbf{T}_1 & \mathbf{T}_2 & \mathbf{T}_3 \end{bmatrix}$
				\ForAll{$\mathbf{A}_{i}$}
					\ForAll{$\mathbf{B}_{j}$}
						\State $ T_{ij} = \mathbf{A}_{i}(\xi_0) \cdot\mathbf{B}_{j}(\xi_0)$
					\EndFor
				\EndFor
			\EndProcedure
			\Statex
			\Procedure{Find local basis system $\mathbf{B}(\xi_r)$}{}			
				\State \textit{1. Get an initial guess on the sweep path }
				\State Get closest point on curve $\mathbf{C}(\xi^{j_0}_r)$
				\State Store $\xi^{j_0}_r$ as initial value for Newton iteration
				\Statex
				\State \textit{2. Apply Newton's method to find $\mathbf{B}(\xi_r)$ (see \cref{fig:RotateBaseFinding}).}
				\State Let $\text{d}\xi$ be a sufficiently small parameter increment for finite differences
				\State Initialize z-coordinate of (first) mapped point of interest $\vert\tilde{P}^1_z \vert \overset{!}{>} \varepsilon$
				\While{$\vert\tilde{P}^1_z \vert > \varepsilon$}
					\State \textit{2.1 Create local basis of sketch $\mathbf{B}(\xi^j_r)$ at $\xi^j_r$}
					\State Create local basis system of the sweep path $\mathbf{A}(\xi^{j}_r)$ at $\mathbf{C}(\xi^{j}_r)$ (e.g. Frenet base)
					\ForAll{$ \mathbf{B}_{i}(\xi^j_r)$}
						\State Get basis vectors of sketch base $\mathbf{B}_i(\xi^j_r) = \mathbf{A}(\xi^j_r)\: \mathbf{T}_{i}$
					\EndFor			
					\Statex
					\State \textit{2.2 Create local basis of sketch $\mathbf{B}(\xi^j_r + \mathrm{d}\xi)$ at $\xi^j_r+ \mathrm{d}\xi$ }
					\State Create local basis system of the sweep path $\mathbf{A}(\xi^{j}_r+ \mathrm{d}\xi)$ at $\mathbf{C}(\xi^{j}_r+ \mathrm{d}\xi)$ (e.g. Frenet base)
					\ForAll{$ \mathbf{B}_{i}(\xi^j_r+ \mathrm{d}\xi)$}
						\State Get basis vectors of sketch base $\mathbf{B}_i(\xi^j_r+ \mathrm{d}\xi) = \mathbf{A}(\xi^j_r+ \mathrm{d}\xi)\: \mathbf{T}_{i}$
					\EndFor			
					\Statex
					\State \textit{2.3 Map point of interest $\mathbf{P}$ to local sketch basis systems $\mathbf{B}(\xi^j_r)$ and $\mathbf{B}(\xi^j_r + \mathrm{d}\xi)$.}
					\State First mapped point $\mathbf{\tilde{P}}^1 = \mathbf{P}\rightarrow \mathbf{B}(\xi^j_r)$
					\State Second mapped point $\mathbf{\tilde{P}}^2 = \mathbf{P}\rightarrow \mathbf{B}(\xi^j_r+ \mathrm{d}\xi)$
					\Statex
					\State \textit{2.4 Next Newton step}
					\State Newton step $\xi^{j+1}_r = \xi^{j}_r - \tilde{P}^1_z\:/\:[(\tilde{P}^2_z-\tilde{P}^1_z)\:/\:\mathrm{d}\xi]$
				\EndWhile
				\Statex
				\State \textbf{return} $\mathbf{B}(\xi^{j_{end}}_r)$
			\EndProcedure
	\end{algorithmic}
\end{algorithm}

\begin{figure}[H]\centering
	\includegraphics[width=14cm]{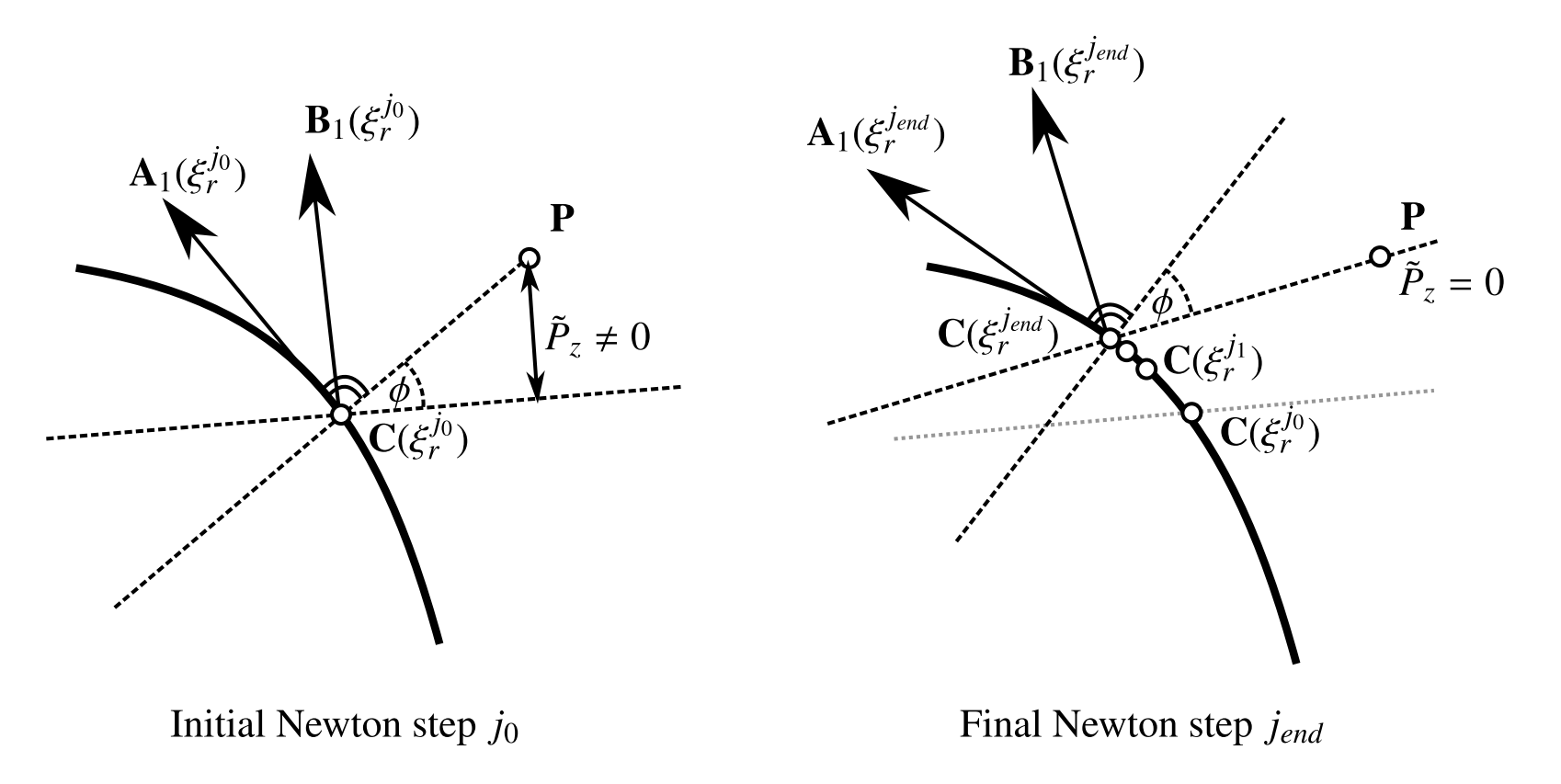}
	\caption{ For the case that the dihedral angle $\phi$  between the local path basis $\mathbf{A}$ and the local sketch basis $\mathbf{B}$ remains constant: Find the point on the curve $\mathbf{C}(\xi_r)$ so that $\tilde{P}_z = 0$ with $\mathbf{\tilde{P}}$ being the point of interest mapped to local sketch basis $\mathbf{B}$. Starting with closest point on the curve $\mathbf{C}(\xi^{j_0}_r)$, a Newton iteration is carried out until  $\tilde{P}_z = 0$.}	
	\label{fig:RotateBaseFinding}	
\end{figure}

Case (b): According to case (a) (\Cref{alg:fixedDihedral}), a local basis system $\mathbf{B}(\xi_r)$ must be found such that $\tilde{P}_z = 0$. In contrast to (a), the basis $\mathbf{B}(\xi_r)$ can be evaluated easier, as the base vectors $\mathbf{B}_i(\xi)$ are constant along the sweep path. Only the origin moves corresponding to the curve point $\mathbf{C}(\xi^j_r)$.

Case (c): As the relations between the local basis system of the path $\mathbf{A}$ and sketch $\mathbf{B}$ are only known at the starting and ending points $\mathbf{C}(\xi_0)$ and $\mathbf{C}(\xi_{end})$, a transformation matrix for both is set-up, $\mathbf{T}_{0}$ and $\mathbf{T}_{end}$, similar to case (a)(\Cref{alg:fixedDihedral}). At each point on the path $\mathbf{C}(\xi_i)$  with both transformation matrices $\mathbf{T}_{0}$ and $\mathbf{T}_{end}$, a local sketch basis is formed $\mathbf{B}^{\mathbf{T}_0}(\xi_i)$ and $\mathbf{B}^{\mathbf{T}_{end}}(\xi_i)$. The basis vectors $\mathbf{B}_i(\xi_i)$ are (linearly) interpolated between  $\mathbf{B}_i^{\mathbf{T}_0}(\xi_i)$ and $\mathbf{B}_i^{\mathbf{T}_{end}}(\xi_i)$ using the arc-length of the current point, similar to the point-in-membership test for lofts (see \cref{sec:pimLoft}).

\newpage
\subsection{ 2D Ray casting on spline curves } \label{sec:2dRayCast}
In the following, an algorithm for ray-casting with splines in 2D is presented. It also deals with the general case in which the ray does not coincide with the positive $x$-axis.

\begin{figure}[H]\centering
	\includegraphics[width=16cm]{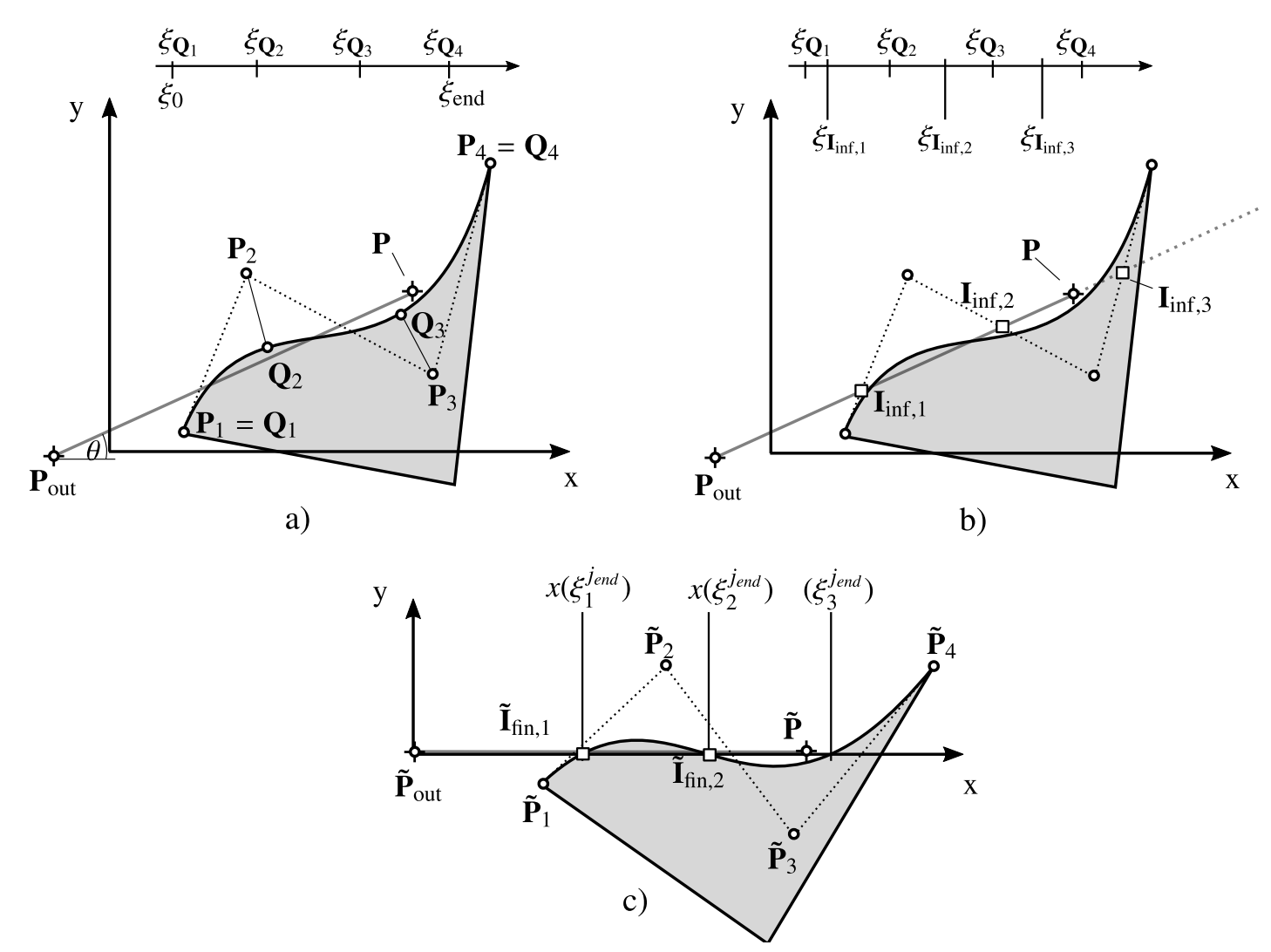}
	\caption{Procedure of ray-casting with splines: a) Mapping the control points to the parameter space $\xi_{\mathbf{Q}_i}$.  \\b) Use linear interpolation to map intersection points with control polygon $\mathbf{I}_{\mathrm{inf},k}$ to spline parameter space $\xi_{\mathbf{I}_{\mathrm{inf},k}}$. c) Perform linear transformation such that the ray lies on the x-axis and perform a zero search to obtain the intersections $\mathbf{\tilde{I}}_{\mathrm{fin},k}$.  }
	\label{fig:RayCastSpline2}	
\end{figure}

\begin{algorithm}
	\caption{Spline ray casting}\label{alg:rayCastSpline}
	\begin{algorithmic}[1]
		\State Let $\mathbf{C}(\xi)$ be a spline defined by its control points $\mathbf{P}_i$
		\State Let $n$ be the number of control points $\mathbf{P}_i$
		\State Let the ray $\textbf{r}_{\mathrm{inf}}$ and the segment $\textbf{r}_{\mathrm{fin}}$ be defined by a point certainly outside $\mathbf{P}_{\mathrm{out}}$ and the point of interest $\mathbf{P}$
		\Statex
		\Procedure{Initialize spline curve}{}
			\State \textit{0. Obtain for each control point a corresponding parameter value (\cref{fig:RayCastSpline2} a)).}			
			\State \textit{Comment: This needs only to be done once at the set-up of the spline}
			\For{$i:=1$ \textbf{to} $n$ } 
				\State Find the closest point $\mathbf{Q}_i$ on the spline to control point $\mathbf{P}_i$ using \cref{eq:inverseMappingSpline}
				\State Find corresponding parameter $\xi_{\mathbf{Q}_i}$
			\EndFor
		\EndProcedure
		\algstore{myalg}
	\end{algorithmic}
\end{algorithm}

\begin{algorithm}[H]
   \ContinuedFloat
   \caption{Spline ray casting (continued)}                     
	\begin{algorithmic} [1]    
		\algrestore{myalg}
		\Procedure{Ray cast on spline}{}
			\State \textit{1. Find intersection points between control polygon and finite and infinite ray (\cref{fig:RayCastSpline2} b)). }
			\State Let $j$ and $k$ denote the $j$-th and $k$-th intersection of the control polygon with the ray segment or the infinite ray respectively.
			\For{$i:=1$ \textbf{to} $n-1$ }
				\State Find intersection point $\mathbf{I}_{\mathrm{fin},j}$ between ray segment $\mathbf{r}_{\mathrm{fin}}$ and control line $\overline{\mathbf{P_i} \mathbf{P_{i+1}} }$
				\State Find intersection point $\mathbf{I}_{\mathrm{inf},k}$ between infinite ray $\mathbf{r}_{\mathrm{inf}}$ and control line $\overline{\mathbf{P_i} \mathbf{P_{i+1}} }$
			\EndFor
			\State Find intersection point $\mathbf{I}_{\mathrm{inf},k}$ between infinite ray $\mathbf{r}_{\mathrm{inf}}$ and line segment $\overline{\mathbf{P_1} \mathbf{P_{n}} }$
			%\Statex
			\State \textit{ 2. Check for number of intersections for finite and infinite ray.}
			\If{$\#\left(\mathbf{I}_{\mathrm{fin}} \right)$ \textbf{equals} $\#\left( \mathbf{I}_{\mathrm{inf}} \right)$} 
				\State \textit{ 2.1 Case: Same number of intersections for finite and infinite ray (\cref{fig:RayCastSpline} a) and b)).}
				%\Statex
				\State \textbf{return}  $modulus \left( \frac{\#\left(\mathbf{I}_{\mathrm{inf}}\right)}{2} \right) $ 
				
			\Else
				\State \textit{ 2.2 Case: Different number of intersections for finite and infinite ray (\cref{fig:RayCastSpline} c) and d)).}

				%\Statex
				\State \textit{3. Interpolate starting values for intersection search (\cref{fig:RayCastSpline2} b)).}
				\ForAll{$\mathbf{I}_{\mathrm{inf},k} $}
					\State Get corresponding control line $\overline{\mathbf{P_i} \mathbf{P_{i+1}} }$
					\State Make a linear interpolation between $\xi_{\mathbf{Q}_i}$ and  $\xi_{\mathbf{Q}_{i+1}}$ to get $\xi_{\mathbf{I}_{\mathrm{inf},k}}$
				\EndFor
				%\Statex
				\State \textit{4. Transform spline and ray to x-axis (\cref{fig:RayCastSpline2} c)).}
				\State Let $\theta$ be the angle between $\mathbf{r}_{\mathrm{inf}}$ and the x-axis
				\State Get transformed ray $\mathbf{\tilde{r}} = \overline{\mathbf{\tilde{P}}_{\mathrm{out}} \mathbf{\tilde{P}}}$ with $\mathbf{\tilde{P}}_{\mathrm{out}} = \mathbf{0}$ and $\mathbf{\tilde{P}} = (\mathbf{P} - \mathbf{P}_{\mathrm{out}} ) \mathbf{T}_{\mathrm{rot}}(\theta)$
				\ForAll{$\mathbf{P}_{i} $}
					\State Get transformed control points $\mathbf{\tilde{P}}_{i} = (\mathbf{P}_{i} - \mathbf{P}_{\mathrm{out}} ) \mathbf{T}_{\mathrm{rot}}(\theta)$
				\EndFor
				\State \textit{5. Newton's method to find zeros of transformed spline (\cref{fig:RayCastSpline2} c)) .}								
				\For{$k:=1$ \textbf{to} $\#\left(\xi_{\mathbf{I}_{\mathrm{inf},k}}\right)$}
					\State Starting value $\xi_k^{j_0} = \xi_{\mathbf{I}_{\mathrm{inf},k}}$
					\State Initialize $\vert y_k^{j_0} \vert \overset{!}{>} \varepsilon$ 
					\While {$\vert y_k^j \vert > \varepsilon$}
						\State Get point on transformed spline $\mathbf{\tilde{C}}(\xi_k^j) = \begin{pmatrix}
					x_k^j \\ y_k^j	\end{pmatrix}$.
						\State Get tangent vector on transformed spline $\mathbf{\dot{\tilde{C}}}(\xi_k^j)$.
						\State Newton step $\xi_k^{j+1} = \xi_k^{j} -  \mathbf{\tilde{C}}(\xi_k^j) / \mathbf{\dot{\tilde{C}}}(\xi_k^j)$
					\EndWhile
					\Statex	
					\If {$x^{j_{end}}_k < \tilde{P}_{x}$}
						\State Append intersection point of transformed spline and finite ray $\mathbf{\tilde{I}}_{\mathrm{fin},k} = \begin{pmatrix}
					x_k^{j_{end}} \\ y_k^{j_{end}}	\end{pmatrix}$.
					\EndIf
				
				\EndFor
				%\Statex	
				\State \textbf{return} $modulus \left( \frac{\#\left( \mathbf{I}_{\mathrm{fin}}\right))}{2} \right) $ 		
			\EndIf
		\EndProcedure
	\end{algorithmic}
\end{algorithm}

%% file: main.bbl
\begin{thebibliography}{10}

\bibitem{Foley1997}
J.~D. Foley, A.~V. Dam, S.~K. Feiner, J.~F. Hughes, and R.~L. Phillips, {\em
  Introduction to {{Computer Graphics}}}.
\newblock {Addison-Wesley}, 1997.

\bibitem{Gomes1991}
A.~J.~P. Gomes and J.~G. Teixeira, ``Form feature modelling in a hybrid
  {{CSG}}/{{BRep}} scheme,'' {\em Computers \& Graphics}, vol.~15, no.~2,
  pp.~217--229, 1991.

\bibitem{Shah1995}
J.~J. Shah and M.~M{\"a}ntyl{\"a}, {\em Parametric and {{Feature}}-{{Based
  CAD}}/{{CAM}}: {{Concepts}}, {{Techniques}}, and {{Applications}}}.
\newblock {John Wiley \& Sons}, 1995.

\bibitem{Cottrell2009}
J.~Cottrell, T.~J. Hughes, and Y.~Bazilevs, {\em Isogeometric {{Analysis}}:
  {{Toward Integration}} of {{CAD}} and {{FEA}}}.
\newblock New York: {Wiley and Sons}, 2009.

\bibitem{Hughes2005}
T.~J.~R. Hughes, J.~A. Cottrell, and Y.~Bazilevs, ``Isogeometric analysis:
  {{CAD}}, finite elements, {{NURBS}}, exact geometry and mesh refinement,''
  {\em Computer Methods in Applied Mechanics and Engineering}, vol.~194,
  pp.~4135--4195, Oct. 2005.

\bibitem{Piegl1997}
L.~Piegl and W.~Tiller, {\em The {{NURBS Book}}}.
\newblock Monographs in Visual Communication, Berlin, Heidelberg: {Springer
  Berlin Heidelberg}, 1997.

\bibitem{Bazilevs2010a}
Y.~Bazilevs, V.~M. Calo, J.~A. Cottrell, J.~A. Evans, T.~J.~R. Hughes,
  S.~Lipton, M.~A. Scott, and T.~W. Sederberg, ``Isogeometric analysis using
  {{T}}-splines,'' {\em Computer Methods in Applied Mechanics and Engineering},
  vol.~199, pp.~229--263, Jan. 2010.

\bibitem{Massarwi2016}
F.~Massarwi and G.~Elber, ``A {{B}}-spline based framework for volumetric
  object modeling,'' {\em Computer-Aided Design}, vol.~78, pp.~36--47, Sept.
  2016.

\bibitem{Zuo2015}
B.-Q. Zuo, Z.-D. Huang, Y.-W. Wang, and Z.-J. Wu, ``Isogeometric analysis for
  {{CSG}} models,'' {\em Computer Methods in Applied Mechanics and
  Engineering}, vol.~285, pp.~102--124, Mar. 2015.
\newblock 00000.

\bibitem{Patera1988}
A.~Patera, ``Nonconforming {{Mortar Elements Methods}}: {{Application}} to
  {{Spectral Dicretizations}}.''
  \url{http://ntrs.nasa.gov/archive/nasa/casi.ntrs.nasa.gov/19890002965.pdf},
  1988.

\bibitem{Natekar2004}
D.~Natekar, X.~Zhang, and G.~Subbarayan, ``Constructive solid analysis: A
  hierarchical, geometry-based meshless analysis procedure for integrated
  design and analysis,'' {\em Computer-Aided Design}, vol.~36, pp.~473--486,
  Apr. 2004.

\bibitem{Schillinger2012a}
D.~Schillinger, L.~Ded{\`e}, M.~A. Scott, J.~A. Evans, M.~J. Borden, E.~Rank,
  and T.~J. Hughes, ``An isogeometric design-through-analysis methodology based
  on adaptive hierarchical refinement of {{NURBS}}, immersed boundary methods,
  and {{T}}-spline {{CAD}} surfaces,'' {\em Computer Methods in Applied
  Mechanics and Engineering}, vol.~249-252, pp.~116--150, 2012.

\bibitem{Parvizian2011}
J.~Parvizian, A.~D{\"u}ster, and E.~Rank, ``Topology optimization using the
  finite cell method,'' {\em Optimization and Engineering}, vol.~13,
  pp.~57--78, July 2011.

\bibitem{Rank2012}
E.~Rank, M.~Ruess, S.~Kollmannsberger, D.~Schillinger, and A.~D{\"u}ster,
  ``Geometric modeling, isogeometric analysis and the finite cell method,''
  {\em Computer Methods in Applied Mechanics and Engineering}, vol.~249-252,
  pp.~104--115, Dec. 2012.

\bibitem{Bungartz2004}
H.-J. Bungartz, M.~Griebel, and C.~Zenger, {\em Introduction to {{Computer
  Graphics}}}.
\newblock {Charles River Media}, 2004.

\bibitem{Requicha1977}
A.~A.~G. Requicha and H.~B. Voelker, {\em Constructive {{Solid Geometry}},
  {{TM}}-25}.
\newblock {Production Automation Project, University of Rochester}, 1977.

\bibitem{Lorensen1987}
W.~E. Lorensen and H.~E. Cline, ``Marching cubes: {{A}} high resolution {{3D}}
  surface construction algorithm,'' in {\em Proceedings of the 14th Annual
  Conference on {{Computer}} Graphics and Interactive Techniques}, (New York,
  NY), pp.~163--169, {ACM Press}, 1987.

\bibitem{Duster2008}
A.~D{\"u}ster, J.~Parvizian, Z.~Yang, and E.~Rank, ``The finite cell method for
  three-dimensional problems of solid mechanics,'' {\em Computer Methods in
  Applied Mechanics and Engineering}, vol.~197, pp.~3768--3782, Aug. 2008.

\bibitem{Hughes2000}
T.~J.~R. Hughes, {\em The Finite Element Method: Linear Static and Dynamic
  Finite Element Analysis}.
\newblock Mineola, NY: {Dover Publications}, 2000.

\bibitem{Reddy1997}
B.~D. Reddy, {\em Introductory {{Functional Analysis}}: {{With Applications}}
  to {{Boundary Value Problems}} and {{Finite Elements}}}.
\newblock No.~27 in Texts in applied mathematics, New York: {Springer},
  1998~ed., Nov. 1997.

\bibitem{Hughes1978}
T.~J.~R. Hughes, ``A simple scheme for developing `upwind' finite elements,''
  {\em International Journal for Numerical Methods in Engineering}, vol.~12,
  pp.~1359--1365, Jan. 1978.
\newblock 00258.

\bibitem{Parvizian2007}
J.~Parvizian, A.~D{\"u}ster, and E.~Rank, ``Finite cell method,'' {\em
  Computational Mechanics}, vol.~41, pp.~121--133, Apr. 2007.

\bibitem{Dauge2015a}
M.~Dauge, A.~D{\"u}ster, and E.~Rank, ``Theoretical and {{Numerical
  Investigation}} of the {{Finite Cell Method}},'' {\em J. Sci. Comput.},
  vol.~65, pp.~1039--1064, Dec. 2015.

\bibitem{Szabo2004}
B.~A. Szab{\'o}, A.~D{\"u}ster, and E.~Rank, ``The p-version of the finite
  element method,'' in {\em Encyclopedia of {{Computational}} Mechanics}
  (E.~Stein, ed.), Chichester, West Sussex: {John Wiley \& Sons}, 2004.

\bibitem{Schillinger2011}
D.~Schillinger and E.~Rank, ``An unfitted hp-adaptive finite element method
  based on hierarchical {{B}}-splines for interface problems of complex
  geometry,'' {\em Computer Methods in Applied Mechanics and Engineering},
  vol.~200, pp.~3358--3380, Nov. 2011.

\bibitem{Schillinger2012b}
D.~Schillinger, M.~Ruess, N.~Zander, Y.~Bazilevs, A.~D{\"u}ster, and E.~Rank,
  ``Small and large deformation analysis with the p- and {{B}}-spline versions
  of the {{Finite Cell Method}},'' {\em Computational Mechanics}, vol.~50,
  pp.~445--478, Feb. 2012.

\bibitem{Duczek2013}
S.~Duczek, M.~Joulaian, A.~D{\"u}ster, and U.~Gabbert, ``Simulation of {{Lamb}}
  waves using the spectral cell method,'' pp.~86951U--86951U--11, Apr. 2013.

\bibitem{Joulaian2013}
M.~Joulaian and A.~D{\"u}ster, ``Local enrichment of the finite cell method for
  problems with material interfaces,'' {\em Computational Mechanics}, vol.~52,
  pp.~741--762, Oct. 2013.

\bibitem{Abedian2013}
A.~Abedian, J.~Parvizian, A.~D{\"u}ster, H.~Khademyzadeh, and E.~Rank,
  ``Performance of {{Different Integration Schemes}} in {{Facing
  Discontinuities}} in the {{Finite Cell Method}},'' {\em International Journal
  of Computational Methods}, vol.~10, p.~1350002, June 2013.
\newblock 00027.

\bibitem{Kudela2016}
L.~Kudela, N.~Zander, S.~Kollmannsberger, and E.~Rank, ``Smart octrees:
  {{Accurately}} integrating discontinuous functions in {{3D}},'' {\em Computer
  Methods in Applied Mechanics and Engineering}, vol.~306, pp.~406--426, July
  2016.

\bibitem{Kollmannsberger2015}
S.~Kollmannsberger, A.~{\"O}zcan, J.~Baiges, M.~Ruess, E.~Rank, and A.~Reali,
  ``Parameter-free, weak imposition of {{Dirichlet}} boundary conditions and
  coupling of trimmed and non-conforming patches,'' {\em International Journal
  for Numerical Methods in Engineering}, vol.~101, pp.~670--699, Mar. 2015.
\newblock 00001.

\bibitem{Ruess2013}
M.~Ruess, D.~Schillinger, Y.~Bazilevs, V.~Varduhn, and E.~Rank, ``Weakly
  enforced essential boundary conditions for {{NURBS}}-embedded and trimmed
  {{NURBS}} geometries on the basis of the finite cell method,'' {\em
  International Journal for Numerical Methods in Engineering}, vol.~95,
  pp.~811--846, Sept. 2013.

\bibitem{Whitney1937}
H.~Whitney, ``On regular closed curves in the plane,'' {\em Compositio
  Mathematica}, vol.~4, pp.~276--284, 1937.

\bibitem{Machchhar2016}
J.~Machchhar and G.~Elber, ``Revisiting the problem of zeros of univariate
  scalar {{B{\'e}ziers}},'' {\em Computer Aided Geometric Design}, vol.~43,
  pp.~16--26, Mar. 2016.

\bibitem{Zander2015}
N.~Zander, T.~Bog, S.~Kollmannsberger, D.~Schillinger, and E.~Rank,
  ``Multi-level hp-adaptivity: High-order mesh adaptivity without the
  difficulties of constraining hanging nodes,'' {\em Computational Mechanics},
  vol.~55, pp.~499--517, Feb. 2015.

\bibitem{Schoberl2003}
J.~Sch{\"o}berl, ``{{NETGEN}}.'' \url{http://www.hpfem.jku.at/netgen/}, 2003.

\end{thebibliography}
